\documentclass[useAMS,usenatbib]{mnras}
\usepackage{graphicx} 
\usepackage{subcaption}
\usepackage{amsmath} 
\usepackage{xcolor}
\usepackage{float}
\usepackage{caption}
\usepackage{placeins}
\usepackage{lipsum}
\usepackage{multicol}
\usepackage{soul}
\usepackage{amssymb}
\usepackage{ragged2e}
\usepackage{threeparttable}
\usepackage{comment}
\def\refjnl#1{{\rm#1}}
\def\aj{\refjnl{AJ}}                   
\def\araa{\refjnl{ARA\&A}}             
\def\apj{\refjnl{ApJ}}                 
\def\apjl{\refjnl{ApJ}}                
\def\apjs{\refjnl{ApJS}}               
\def\ao{\refjnl{Appl.~Opt.}}           
\def\aap{\refjnl{A\&A}}                
\def\aapr{\refjnl{A\&A~Rev.}}          
\def\mnras{\refjnl{MNRAS}}             
\def\pasj{\refjnl{PASJ}}               
\def\physrep{\refjnl{Phys.~Rep.}}   

\def\be{\begin{equation}} 
\def\ee{\end{equation}} 
\def \dd{\mathrm{d}}

\def\kms{\,{\rm {km\, s^{-1}}}}

\def\gsim{\lower.5ex\hbox{\gtsima}} 
\def\lsim{\lower.5ex\hbox{\ltsima}} \def\gtsima{$\; \buildrel > \over 
\sim \;$} \def\ltsima{$\; \buildrel < \over \sim \;$} \def\prosima{$\; 
\buildrel \propto \over \sim \;$} \def\gsim{\lower.5ex\hbox{\gtsima}} 
\def\lsim{\lower.5ex\hbox{\ltsima}} 
\def\simgt{\lower.5ex\hbox{\gtsima}} 
\def\simlt{\lower.5ex\hbox{\ltsima}} 
\def\simpr{\lower.5ex\hbox{\prosima}} \def\la{\lsim} \def\ga{\gsim} 
  
 \def\gtsima{$\; \buildrel > \over \sim \;$} 
\def\ltsima{$\; \buildrel < \over \sim \;$} 
\def\gsim{\lower.5ex\hbox{\gtsima}} 
\def\lsim{\lower.5ex\hbox{\ltsima}} 
\def\simgt{\lower.5ex\hbox{\gtsima}} 
\def\simlt{\lower.5ex\hbox{\ltsima}} 
\def\simpr{\lower.5ex\hbox{\prosima}} 
\def\la{\lsim} 
\def\ga{\gsim}

\def\msun{\,{\rm \Msun}}

\def\E3{{\cal E}_{\rm g}^{III}}

\def\Msun{\rm M_\odot}
\def\lsun{\rm L_\odot}

\def\Msun{\rm M_\odot}

\def\M*{M_*}

\def\Z*{Z_*}
\def\L*{L_*}
\def\muv{\rm M_{UV}}

\def\fej{f_*^{\rm{ej}}}
\def\feff{f_*^{\rm{eff}}}

\newcommand\code[1]{\textsc{\MakeLowercase{#1}}}

\newcommand{\angstrom}{\mbox{\normalfont\AA}}
\def \sixteenth{$16^{\rm{th}} \,$}
\def \eightyfourth{$84^{\rm{th}} \,$}
\def \fescuv{f_{\rm{esc}}^{\rm{UV}}}

\title[Dust in high-z galaxies]{The dust enrichment of early galaxies in the JWST and ALMA era }
\author[Mauerhofer \& Dayal]{Valentin Mauerhofer$^{1}$\thanks{v.mauerhofer@rug.nl} and Pratika Dayal$^1$\\ 
$^{{1}}$ Kapteyn Astronomical Institute, University of Groningen, PO Box 800, 9700 AV Groningen, The Netherlands\\}

\begin{document}
\maketitle
\label{firstpage}
\pagerange{\pageref{firstpage}--\pageref{lastpage}}

\begin{abstract}

Recent observations with the James Webb Space Telescope are yielding tantalizing hints of an early population of massive, bright galaxies at $z > 10$, with Atacama Large Millimeter Array (ALMA) observations indicating significant dust masses as early as $z\sim 7$. To understand the implications of these observations, we use the \code{delphi} semi-analytic model that jointly tracks the assembly of dark matter halos and their baryons, including the key processes of dust enrichment. Our model employs only two redshift- and mass-independent free parameters (the maximum star-formation efficiency and the fraction of supernova energy that couples to gas) that are tuned against all available galaxy data at $z \sim 5-9$ before it is used to make predictions up to $z \sim 20$. Our key results are: {\it (i)} the model under-predicts the observed ultraviolet luminosity function (UV LF) at $z > 12$; observations at $z>16$ lie close to, or even above, a ``maximal" model where all available gas is turned into stars; {\it (ii)} UV selection would miss 34\% of the star formation rate density at $z \sim 5$, decreasing to 17\% by $z \sim 10$ for bright galaxies with $\rm{M_{UV}} < -19$; {\it (iii)} the dust mass ($M_d$) evolves with the stellar mass ($M_*$) and redshift as $\log(M_d) = 1.194\log(M_*) + 0.0975z - 5.433$; {\it (iv)} the dust temperature increases with stellar mass, ranging between $30-33$ K for $M_* \sim 10^{9-11}M_\odot$ galaxies at $z \sim 7$. Finally, we predict the far infrared LF at $z \sim 5-20$, testable with ALMA observations, and caution that spectroscopic redshifts and dust masses must be pinned down before invoking unphysical extrema in galaxy formation models.

\end{abstract}


\section{Introduction}

The first billion years after the Big Bang saw the emergence of the first galaxies, whose stellar populations created the first heavy elements and dust \citep[for a review see e.g.][]{Maiolino19} as well as the first hydrogen-ionizing photons that started the process of cosmic reionization \citep[for a review see e.g.][]{dayal2018}. The emergence of these first systems and their large-scale effects remain key outstanding questions in our cosmic timeline. 

Over the past decade, tremendous efforts have been made to build a global picture of galaxy formation and evolution at high-redshifts, through a combination of multi-wavelength observations using facilities such as the Hubble Space Telescope (HST), the Very Large Telescope (VLT) and the Subaru Telescope to name a few \citep[for reviews see e.g.][]{Dunlop13,Stark16}. More recently, the Atacama Large Millimeter Array (ALMA) has started providing unprecedented views of the dust content of early galaxies at redshifts $z \sim 4.4-7.5$ through the ALMA Large Program to INvestigate C+ at Early Times \citep[ALPINE;][]{Dessauges-Zavadsky20, Bethermin20} and the ALMA Reionization Epoch Bright Line Emission Survey \citep[REBELS;][]{REBELS, Inami22}. A key issue in determining the dust masses of early galaxies is that the observed far Infrared (FIR) continuum emission is characterized by key two quantities - the dust temperature ($T_d$) and the dust mass ($M_d$). Unless multi-band dust measurements are available \citep[see e.g.][]{faisst2020, bakx2021}, these two quantities are degenerate, requiring an assumption on the mass-weighted dust temperature in order to infer the associated dust mass. Despite these caveats, a puzzle is the extremely high dust-to-stellar-mass ratios, ranging between $0.012-3\%$, obtained for star forming galaxies with stellar masses $M_* \sim 10^{8.3-10.5}\msun$, at $z \gsim 7$ \citep[e.g.][]{watson2015, laporte2017, hashimoto2019,bakx2020, reuter2020, REBELS}. Further, a key property of dust is its ability to absorb (non-ionizing) ultra-violet (UV) photons that are re-emitted in the FIR \citep[see e.g.][]{dayal2010}. ALMA REBELS observations have recently allowed such FIR luminosity functions (LFs) to be mapped out at $z \sim 7$ \citep{Barrufet23}.

Furthermore, the James Webb Space Telescope (JWST) has recently started providing ground-breaking views of galaxy formation at $z \sim 9-18$, allowing us to reach this unexplored territory of galaxy formation \citep{adams2022, Atek23, Bouwens22jwst, Bradley22, Naidu22}. This has led to estimates of the global UV LF up to $z \sim 18$ although caution must be exerted when using the LF at $z \gsim 12$ where the redshift and nature of the sources remains debated \citep{adams2022, naidu2022, haro2023}. Surprisingly, the UV LF seems to show almost no evolution at the bright end ($\muv \lsim -22$) at $z \sim 4-13$ \citep[e.g.][]{bowler2020,harikane2022a} showing a possible excess in number density when compared to a Schechter function \citep[e.g.][]{bowler2015,Ono18}. This has led a number of explanations including a co-evolution of halo mass function and the dust content of galaxies \citep{Ferrara22b}, UV contribution from black-hole accretion powered active galactic nuclei \citep[AGN;][]{Ono18,Piana22,pacucci2022}, observational biases causing us to observe only exceptionally starbursting galaxies \citep{Mirocha23} or an initial mass function (IMF) that evolves with redshift \citep{pacucci2022,yung2023}. Finally, the JWST has also allowed to further probe the stellar mass function (SMF) out to $z\sim 8$ \citep[e.g.][]{Navarro-Carrera23} despite caveats on the assumed star formation history that can lead to a significant variations in the inferred stellar mass \citep[e.g.][]{Topping22}. 

In view of these recent advances, a number of models have been used to explore the physical mechanisms of dust production and evolution as well as the effects of dust on early galaxy observables. The approaches adopted range from hydrodynamical simulations that model small-scales processes such as dust growth, dust destruction, grain size distribution and the geometry of dust and stars \citep[e.g.][]{Bekki15,Aoyama17,McKinnon18,Parente22,Trebitsch23} to simulations that have been post-processed with dust models to compute the dust content and attenuation \citep[e.g.][]{dayal11,Mancini15,Narayanan18,Wilkins18,Li19,Ma19,Graziani20,Vogelsberger20,Vijayan23} to semi-analytic models \citep[e.g.][]{Popping17,Vijayan19,Triani20,dayal2022} and analytic formalisms \citep[e.g.][]{Ferrara22b, DeRossi23}.

In this work we make use of the broad mass range and flexibility offered by the \code{delphi} semi-analytic model \citep{dayal2014, dayal2022} to study the dust content of high-redshift galaxies, including the effect of dust on their visibility and its emission in the FIR. A key strength of this model is that it only has two mass- and redshift-independent free parameters and is base-lined against all available data-sets at $z \sim 5-9$ before its predictions are extended to even higher redshifts. 

Throughout this paper, we adopt a $\Lambda$CDM model with dark energy, dark matter and baryonic densities in units of the critical density as $\Omega_{\Lambda}= 0.691$, $\Omega_{m}= 0.308$ and $\Omega_{b}= 0.049$, respectively, a Hubble constant $H_0=100\, h\,{\rm km}\,{\rm s}^{-1}\,{\rm Mpc}^{-1}$ with $h=0.67$, spectral index $n=0.96$ and normalisation $\sigma_{8}=0.81$ \citep[][]{planck2016}. Additionally, we use the stellar library BPASSv2.2.1 \citep{BPASS1,BPASS2}. This library assumes a Kroupa IMF \citep{Kroupa01}, with a slope of $-1.3$ between 0.1 and 0.5 $\Msun$ and of $-2.35$ between 0.5 and 100 $\Msun$. Finally, we use comoving units and magnitudes in the standard AB system \citep{ABmag} throughout the paper.

The paper is structured as follows: in Sec. \ref{sec:methods} we detail the \code{Delphi} model, including a description of the halo merger tree, the computation of star-formation, supernovae feedback, dust evolution and the associated luminosities. In Sec. \ref{sec:uv_obs} we present the results of our model in terms of UV observables, such as the LF and the cosmic UV density, as well as the mass-luminosity relation and the stellar mass function. In Sec. \ref{sec:dust_properties}, we detail the derived dust properties of high-redshift galaxies, including the dust mass, dust temperature and UV escape fraction, along with analytical relations between those quantities and the stellar mass and star-formation rate. In Sec. \ref{sec:dust_detectability} we discuss the infrared emission of high-redshift galaxy spectra, and compare our results with far-infrared (FIR) LFs from the literature. Finally, we summarize and discuss our results in Sec. \ref{sec:discussion}.

\section{Theoretical model} \label{sec:methods}
In this section, we briefly describe the theoretical model used to study the assembly of dark matter halos and their baryonic components at $z \sim 4.5-20$; interested readers are referred to our previous papers \citep{dayal2014, dayal2022} for complete details. We start with a description of the merger tree (Sec. \ref{mt}) before discussing the star formation prescription and the associated supernova (SN) feedback (Sec. \ref{sec:sfr-feedback}), the dust enrichment of early galaxies (Sec. \ref{sec:dust_modeling}) and the resulting luminosities in both the UV and FIR (Sec. \ref{sec:UV_IR}).

\subsection{Halo merger tree and gas accretion}
\label{mt}
Starting at $z = 4.5$ we build merger trees for 600 galaxies, up to $z \sim 40$, uniformly distributed in terms of the halo mass (in log space) between $\log(M_h/\Msun) = 8-14$ using the binary merger tree algorithm from \cite{parkinson2008}. We impose a mass resolution of $10^8 \Msun$ and use a constant redshift-step of 30 Myr for the merger tree so that all Type II SN (SNII) explode within a single redshift-step, preventing the need for delayed SN feedback. Each halo is assigned a number density by matching to the Sheth-Tormen \citep{sheth1999} halo mass function (HMF) at $z =4.5$ and this number density is propagated throughout its merger tree. We have confirmed that the resulting HMFs are in accord with the Sheth-Tormen HMFs at all higher redshifts, up to $z \sim 20$.

The first progenitors (``starting leaves") of any merger tree are assigned an initial gas mass that is linked to the halo mass through the cosmological ratio such that  
$M_{\rm{g}}^{\rm i} = (\Omega_b/\Omega_m) M_h$. At every further redshift-step, the total halo mass is determined by the sum of the dark matter mass brought in by mergers and smooth-accretion from the intergalactic medium (IGM). While we assume the accreted gas mass to be proportional to the accreted dark matter mass, the merged gas mass is determined by the gas mass left in the merging progenitors after star formation and the associated SNII feedback.

\subsection{Star formation and supernova feedback} \label{sec:sfr-feedback}
We start by computing the newly formed stellar mass in a given redshift-step as 
\be \label{eq:mstar}
\Delta M_*(z) = f_*^{\rm{eff}} M_{\rm{g}}^{\rm i}(z),
\ee
where $f_*^{\rm{eff}}$ is the effective star formation efficiency and $M_{\rm g }^{\rm i}$ is the (initial) gas mass at the start of the redshift-step. We assume this mass to have formed uniformly over $t_s=30 \rm{Myr}$ to obtain the star formation rate (SFR) $\psi(z)=\Delta M_*(z)/t_s$. The $f_*^{\rm{eff}}$ value for any halo is the minimum between the star formation efficiency that produces enough SNII energy to unbind the remainder of the gas ($\fej$) and a maximum star formation efficiency parameter ($f_*$) i.e. $f_*^{\rm{eff}}=\min(f_*,\fej)$. While galaxies with $f_*^{\rm{eff}} = f_*$ are efficient star-formers, those with $f_*^{\rm{eff}} = \fej$ comprise ``feedback-limited" systems that can unbind all of their gas content due to SN feedback. 

To compute $\fej$, we start by calculating the energy $E_{\rm{ej}}$ required to unbind the gas left after star formation
\be
E_{\rm{ej}} = (M_{\rm{g}}^{\rm i} - \Delta M_*) v_c^2,
\ee
where $v_c$ is the halo rotational velocity. This is compared to the SNII energy
\be
E_{\rm{SN}} = f_w v_s^2 \Delta M_*,
\ee
where $f_w$ is the fraction of SNII energy coupling to the gas and $v_s^2 = \nu E_{51} = 747 \kms$. Here $\nu=0.011$ is the SNII rate for our chosen Kroupa IMF and we assume each SNII to produce $E_{51}= 10^{51}{\rm erg}$ of energy. The parameter $\fej$ is the star-formation efficiency that would result in an equality between $E_{\rm{SN}}$ and $E_{\rm{ej}}$, i.e., 
\be \label{eq:def_fej}
\fej = \frac{v_c^2}{v_c^2 + f_w v_s^2}.
\ee

With this formalism, the ejected gas mass at any step can be calculated as
\be \label{eq:Mej}
M_{\rm{ej}} = \frac{E_{\rm{SN}}}{E_{\rm{ej}}} (M_{\rm{g}}^{\rm i} - \Delta M_*) = \frac{f_w v_s^2}{v_c^2} \Delta M_*.
\ee

We note that while $f_w$ essentially determines the faint-end of the UV LF and the low-mass end of the SMF, $f_*$ is crucial in determining the high-mass end of the SMF and the bright-end of the UV LF. However, the bright end of the UV LF is also shaped by the presence of dust as detailed in the next section. Simultaneously matching to the observed UV LF and SMF at $z\sim 5-9$, including the impact of dust attenuation, requires $f_* = 15\%$ and $f_w = 6\%$ - these are the free parameter values used in the {\it fiducial} model.

\subsection{Dust modeling} \label{sec:dust_modeling}
We briefly describe our dust model here and interested readers are referred to 
\citet{dayal2022} for complete details. We use a coupled set of equations to model the time-evolution of the gas-phase metal ($M_Z$) and dust masses ($M_d$), assuming perfect mixing of gas, metals and dust, such that
\be
\frac{\dd M_{Z}}{\dd t} = \dot{M}_{Z}^{\rm{pro}} - \dot{M}_{Z}^{\rm{eje}} - \dot{M}_{Z}^{\rm{ast}} - \dot{M}_d^{\rm{gro}} + \dot{M}_d^{\rm{des}} \\
\ee
\be
\frac{\dd M_d}{\dd t} = \dot{M}_d^{\rm{pro}}  - \dot{M}_d^{\rm{eje}} - \dot{M}_d^{\rm{ast}} + \dot{M}_d^{\rm{gro}}- \dot{M}_d^{\rm{des}} .
\ee

Starting with metals, the different terms represent the rates of metal production ($\dot M_Z^{\rm{pro}}$) for which we use the mass- and metallicity-dependent stellar yields between $1-50~\Msun$ \citep{Kobayashi20}, ejection in SNII-driven winds ($\dot M_Z^{\rm{eje}}$), astration into star formation ($\dot M_Z^{\rm{ast}}$), metals lost into dust growth in the interstellar medium (ISM; $\dot M_d^{\rm{gro}}$) and the metals returned to the ISM due to dust destruction ($\dot M_d^{\rm{des}}$). As for dust, we assume that it is mostly produced by SNII, with each SNII producing $0.5 \Msun$ of dust \citep{dayal2022}, with asymptotic giant branch stars (AGBs) having a negligible contribution \citep[e.g.][]{dayal2010,lesniewska2019}. The different terms represent the rates of dust production ($\dot{M}_d^{\rm{pro}}$) in SNII, dust destruction in SNII shocks ($\dot{M}_d^{\rm{des}}$), ejection in winds ($\dot{M}_d^{\rm{eje}}$), loss in astration ($\dot{M}_d^{\rm{ast}}$) and increase due to ISM grain growth.

Assuming perfect mixing, the dust (metals) lost in outflows and astration are equal to the dust-to-gas (metal-to-gas) ratio multiplied by the gas mass lost to these processes. Finally, we model ISM grain growth as \citep{Dwek98}
\be 
\dot{M}_d^{\rm{gro}} = X_c \left(1 - \frac{M_d}{M_d + M_{Z}} \right) \frac{M_d}{\tau_{\rm{acc}}},
\ee
where $\tau_{\rm{acc}} = \tau_0 (Z/Z_{\odot})^{-1}$ and $X_c$ is the fraction of cold ISM gas where such grain growth can take place; we use a value of $X_c=0.5$ based on high-resolution simulations of early galaxies \citep[e.g.][]{Pallottini19}. Finally, $\tau_0$ is the dust accretion timescale and $Z/Z_{\odot}$ is the gas-phase metallicity in solar units. Since $\tau_0$ is relatively poorly known, and changing its value from 30 to 0.3 Myr only changes the dust mass by a factor two \citep{dayal2022}, we adopt $\tau_0=30$ Myr as our {\it fiducial} dust grain-growth timescale. 

\subsection{The emerging UV and FIR luminosities} \label{sec:UV_IR}
We start by calculating the intrinsic luminosity (${L_{\rm UV}^{\rm int}}$) at rest-frame $1500 \angstrom$ assuming a continuous star-formation over the 30 Myr redshift-steps of the merger tree and using the stellar metallicity of each stellar population as inputs for the BPASS (v2.2.1) stellar population synthesis model \citep{BPASS1,BPASS2}.

\begin{figure*}
  \resizebox{\hsize}{!}{\includegraphics{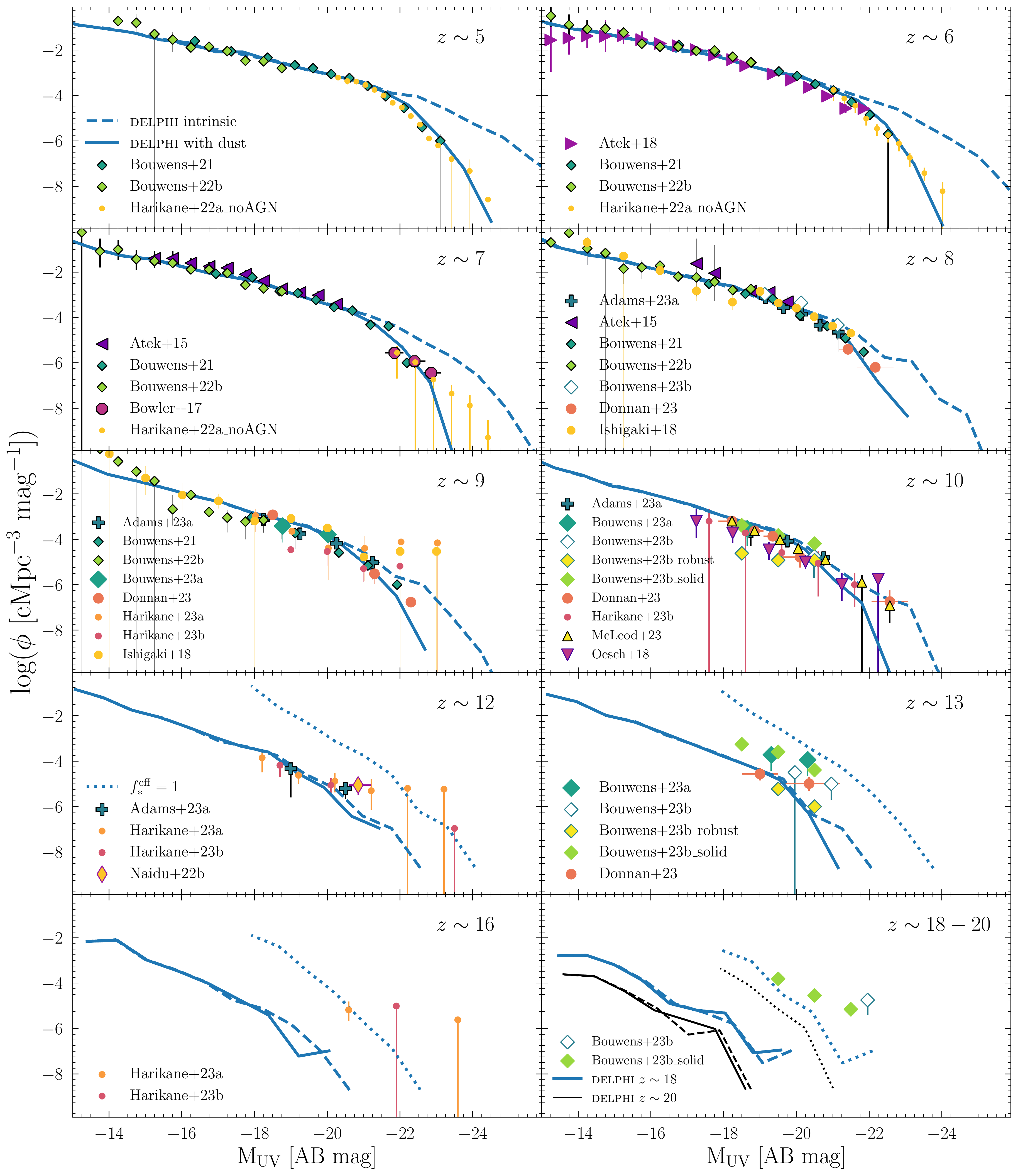}}
  \caption{The UV LF at $z\sim 5-20$, as marked in the panels. In each panel, the dashed and solid lines represent the intrinsic and dust-attenuated UV LFs from the theoretical model, respectively. Finally, the dotted lines at $z \gsim 13$ represent the upper limit to the theoretical UV LF with $f_*=1$ and no feedback. In each panel, points show a compilation of the data from a number of different observational works \citep[including][]{Atek15, Bowler17, Atek18, Ishigaki18, Oesch18, Bouwens21b, Bouwens22faint, Bouwens22jwst, Bouwens23_solid, Naidu22, Adams23, Donnan23, Harikane22a, Harikane23_spec, Harikane23, McLeod23}, as marked.}
  \label{fig:luminosity_function}
\end{figure*}

We then calculate the dust-attenuated ``observed" UV luminosity (${L_{\rm UV}^{\rm obs}}$) as follows \citep[see also][]{dayal2022}: we assume carbonaceous/graphite dust with a single grain size of $a=0.05 \mu m$ and a density $s = 2.25 \rm{g \, cm^{-3}}$ \citep{Todini01,Nozawa03}. We model the dust distribution as a sphere of radius ($r_d$) equal to the gas radius, which is calculated as $r_{\rm{gas}} = 4.5 \lambda r_{\rm{vir}}$ \citep{Ferrara00}. Here, $r_{\rm{vir}}$ is the halo virial radius and the spin parameter is assumed to have an average value of $\lambda = 0.04$ \citep{dayal2018}. Recent ALMA observations \citep{Fujimoto20, Fudamoto22} have shown a gas radius that remains constant between $z \sim 4-7$ for galaxies at a fixed UV luminosity. This is interpreted as gas occupying a larger fraction of the halo volume with increasing redshift. We include this effect by calculating the gas radius as 
\be 
r_d = r_{\rm{gas}} =  4.5 \times 0.04 \left( \frac{1+z}{7} \right) r_{\rm{vir}}.
\ee
This results in a constant radius for a fixed halo mass as a function of redshift. In this slab configuration, the optical depth of the dust is $\tau_d = 3 M_d/(4 \pi r_d^2 a s)$.
The corresponding escape fraction of UV continuum photons is 
\be 
\fescuv = \frac{1 - e^{-\tau_d}}{\tau_d}.
\ee

The dust attenuated UV luminosity is obtained by multiplying the intrinsic UV luminosity by this escape fraction:
\be 
L_{\rm{UV}}^{\rm{obs}}(\lambda) = \fescuv L_{\rm{UV}}^{\rm{int}}(\lambda).
\ee

Concerning the infrared emission, $L_{\rm{FIR}}$, we assume an energy balance between the non-ionizing UV radiation (rest-frame 912-4000\AA) absorbed by dust and the subsequent infrared emission \citep[see e.g.][]{dayal2010}. To compute $L_{\rm{FIR}}$, we first integrate the UV spectra for each source over the wavelength range 912-$4000 \angstrom$ which yields the total FIR luminosity 
\be 
L_{\rm{FIR}} = (1-f_{\rm{esc}}^{\rm{UV}}) \int_{912}^{4000} L_{\rm{UV}}^{\rm{int}}(\lambda) \dd\lambda,
\ee
Finally, the peak of the dust emission temperature, assuming black-body emission, is computed as \citep{dayal2010}:
\be \label{eq:dust_temperature}
T_d = 6.73 \left( \frac{L_{\rm{FIR}} / L_{\odot}}{M_d / M_{\odot}} \right)^{1/(\beta+4)} \rm{K},
\ee
where the dust emissivity index is assumed to have a value of $\beta=2$ \citep{draine1984}. We have calculated this temperature solely based on emission of the galaxy itself. We have ignored the additional heating term is provided by the cosmic microwave background \citep[CMB;][]{dacunha2013} whose temperature scales with redshift as $T(z) = T_0 (1+z)$ K where $T_0 = 2.732$ K. In addition to setting the floor for both the gas and dust temperatures (which becomes increasingly important with increasing redshift), the CMB also provides the background against which dust emission is observed.

\section{The impact of dust on early galaxy observables}
\label{sec:uv_obs}
As a first step, in Sec. \ref{sec_uvlf} we show that our choice of model parameters reproduces the observed UV LF at $z \sim 5-9$ before showing predictions up to $z \sim 20$. We then show the redshift evolution of the cosmic UV luminosity density up to $z \sim 20$, using different magnitude thresholds to compare with observations. This is followed by the relation between both the intrinsic and observed UV magnitudes and the stellar mass in Sec. \ref{sec_msuv} before we show our predictions of the stellar mass function and the corresponding stellar mass density at $z \sim 5-20$ in Sec. \ref{sec_smf}.

\subsection{Redshift evolution of the UV LF}
\label{sec_uvlf}
We begin by showing both the intrinsic and dust-attenuated UV LFs at $z \sim 5-20$ in Fig. \ref{fig:luminosity_function}. We find that, within error bars, the intrinsic UV LF is in good accord with the observed data for $\muv \gsim -21$ at all $z \sim 5-12$. 
This indicates that SNII feedback rather than dust-attenuation plays a dominant role in shaping the observed UV LF for low to intermediate mass galaxies. 
At $z \sim 6-8$, while the theoretical UV LF is in good accord with the results from \citet{Bouwens22faint}, it does not show the slight flattening/downturn seen in the $z \sim 6$ data collected for the faintest (lensed) sources with $\muv \gsim -15$ \citep{Atek18}. This could possibly be attributed to physical effects not considered here, such as the impact of reionization feedback reducing the gas masses (and therefore star forming capabilities) of such low-mass objects \citep[e.g.][]{hutter2021} or the impact of observational uncertainties such as those associated with lensing systematics \citep[e.g.][]{Atek18}. 

\begin{figure}
  \resizebox{\hsize}{!}{\includegraphics{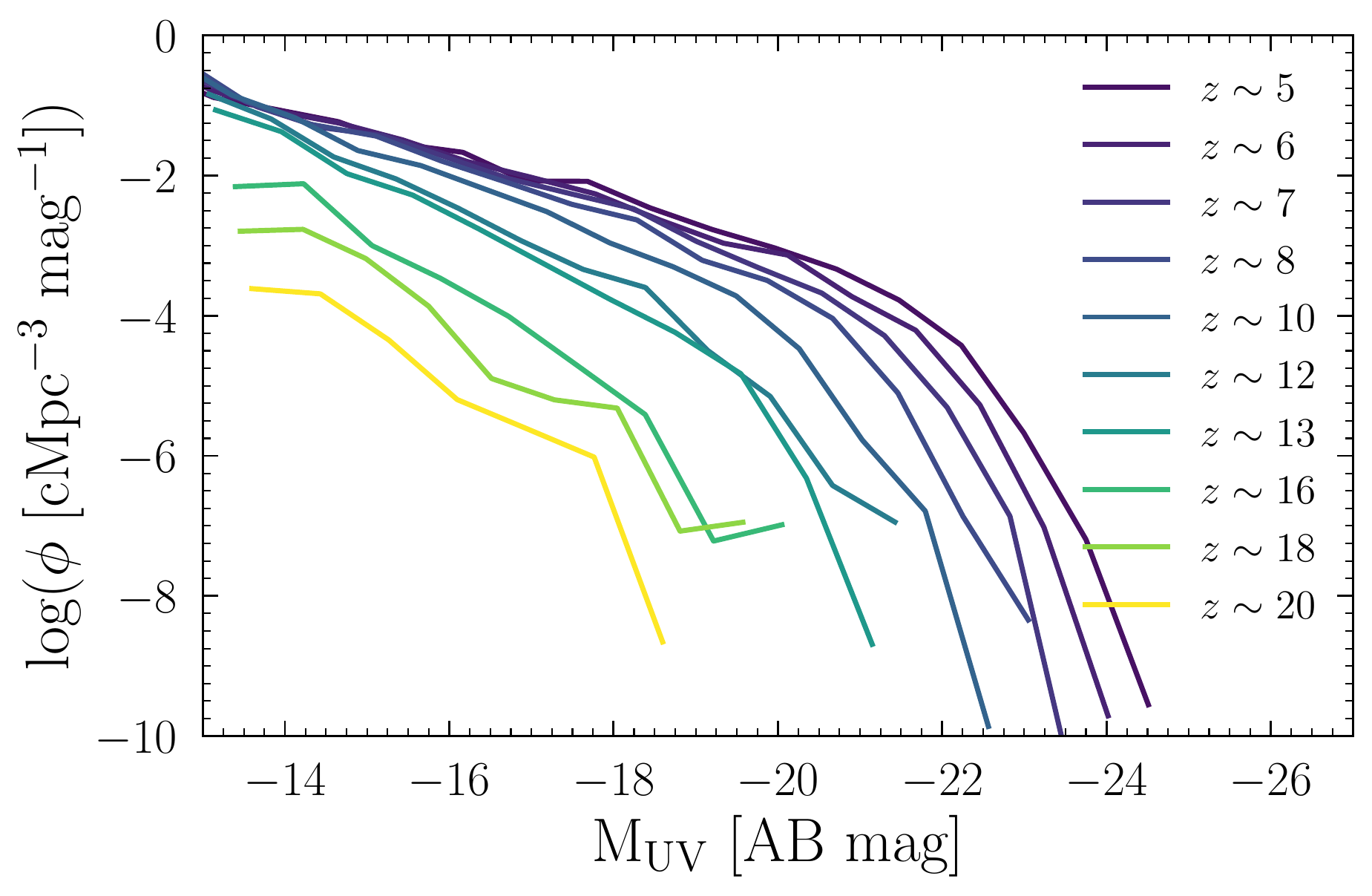}}
  \caption{The redshift evolution of the dust-attenuated UV LF from our model between $z \sim 5-20$, as marked.} 
  \label{fig:UV_LF_all_z}
\end{figure}

Interpreting the bright end of the UV LF is complicated by the fact that, in addition to dust attenuation, black-hole accretion-powered luminosity can have a significant impact on the LF at $\muv \lsim -21$ at $z \sim 5-6$ \citep[e.g.][]{Ono18,kulkarni2019,Piana22}. For this reason, we limit our comparison to the observational UV LF from the star-forming galaxy sample (excluding AGNs) at these redshifts \citep{Harikane22a}. We find that the impact of dust becomes relevant at $\muv \lsim -21$ at $z \sim 5-10$ with dust attenuation playing a negligible role at $z \gsim 12$ where extremely massive, dusty galaxies have not had time to form. Further, while the theoretical dust-attenuated UV LF is in agreement with all observations of the UV LF at $z \sim 5-9$, it under-predicts the number density for the brightest galaxies (with $\muv \sim -22.5$) at $z \sim 10-11$ \citep{Donnan23,McLeod23}. This could be explained by e.g. radiative pressure ejecting dust from such systems which have high specific star formation rates \citep{Ferrara22b} or the dust radius being even larger. We also caution that our homogeneous dust distribution model misses crucial effects such as dust being either clumped/spatially segregated from star forming regions as indicated by REBELS observations \citep[e.g.][]{dayal2022,Inami22} which could have implications for the UV-visibility of these early galaxies.

At $z \gsim 12$, while our UV LF matches to the observations for $\muv \gsim -20$, we under-predict the number density for brighter sources observed by a number of works \citep[e.g.][]{Adams23,Bouwens22jwst,Bouwens23_solid,Naidu22,Donnan23} which increases to an under-prediction by around three orders of magnitude at $z \sim 16-18$ \citep[comparing to][]{Harikane23,Bouwens23_solid}. Although spectroscopic confirmations are crucial in validating the high-redshift nature of these sources, theoretically, such high number densities could be explained by these galaxies being extreme star-formers that significantly lie above the average star formation rate-halo mass relation \citep[e.g.][]{Harikane22a,pacucci2022,Mason23} or having a more top-heavy compared to the Kroupa IMF used here \citep[e.g.][]{pacucci2022,yung2023}.

As a sanity check, we also calculate the ``maximal" UV LF allowed by our model at $z \gsim 12$, assuming no feedback and a star formation efficiency of $\feff=1.0$. Although this extreme model lies above the observations at $z \sim 12-13$ and matches the data at $z \sim 16$ \citep[from][]{Harikane23}, it is still about 0.5 dex below the highest-redshift observations at $z \sim 18$ \citep[from][]{Bouwens22jwst,Bouwens23_solid}. We however caution that spectroscopic confirmations are crucial to validate these ultra-high redshifts \citep[e.g.][]{adams2022, naidu2022, haro2023} before theoretical models are pushed to their extreme limits.

Finally, for clearer visualization, we show the redshift evolution of the dust-obscured UV LF between $z \sim 5-20$ in Fig. \ref{fig:UV_LF_all_z}. At the faint end ($\muv \sim -13$), the amplitude of the observed UV LF is almost constant between $z \sim 5-13$ and shows the expected decline with increasing luminosity. For example, the number of systems with $\muv \sim -18$ falls by about two orders of magnitude between $z \sim 5$ and 13. As expected, we probe to increasingly higher luminosities with decreasing redshifts as more and more massive systems assemble, with the LF extending to $\muv \sim -24.5$ ($\sim -21$) at $z \sim 5$ (13). At $z \sim 16$, the amplitude of the UV LF drops rapidly at all luminosities due to a combination of the evolution of the HMF and such low-mass halos being feedback-dominated. Finally, we note that despite the inclusion of dust, our results do not show the lack of evolution of the bright end of the UV LF seen in observations at $z \sim 5-13$ \citep[e.g.][]{Harikane23,bowler2020}; a part of this could be attributed to the increasing contributions of AGN at the bright-end with decreasing redshift.

\begin{figure}
  \resizebox{0.5\textwidth}{!}{\includegraphics{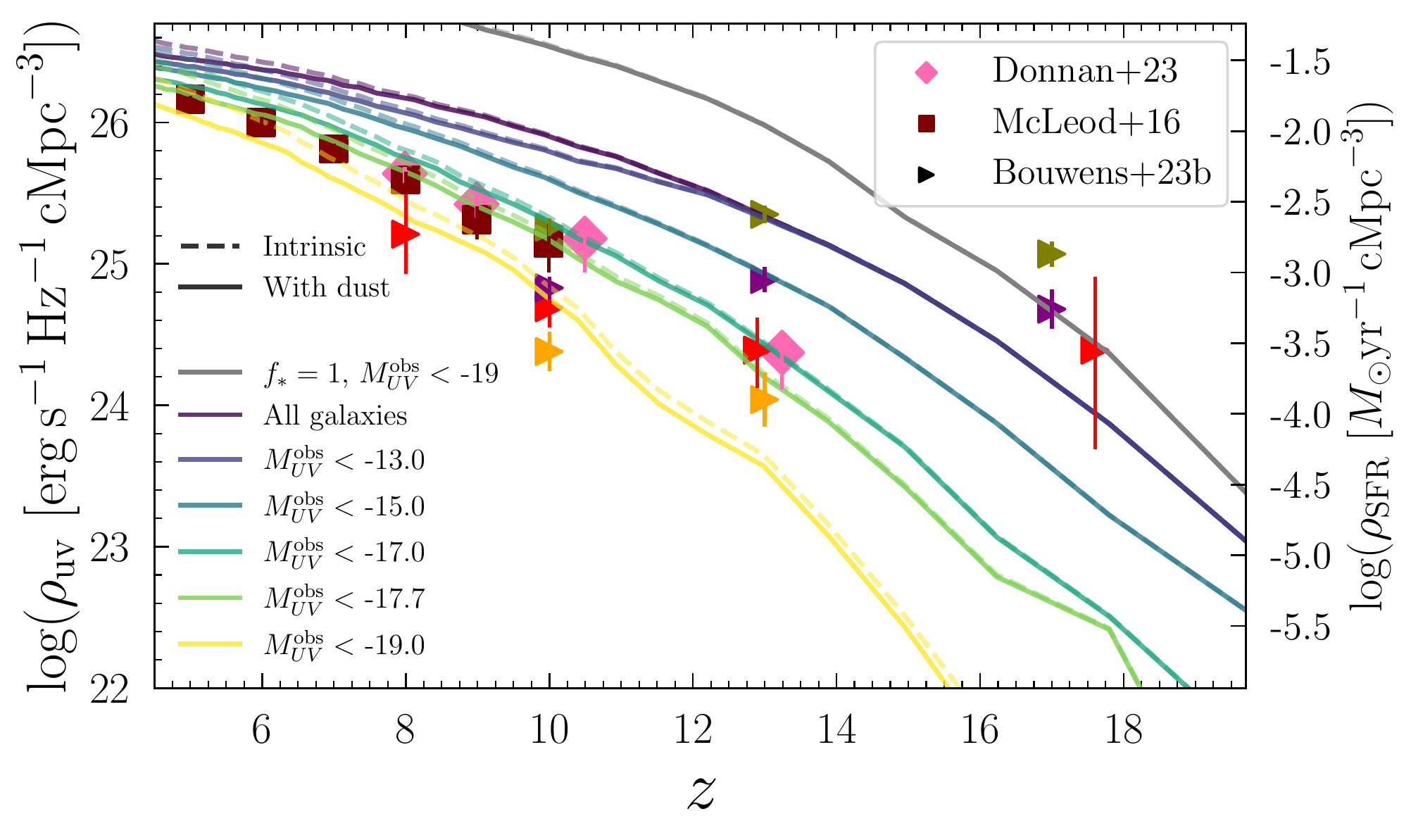}}

  \caption{The redshift evolution of the UV luminosity density between $z \sim 5-20$. We also show the corresponding star formation rate density using a conversion factor between the star formation rate and UV luminosity of $\kappa_{\rm UV} = 1.15 \times 10^{-28}$ [$\rm{M}_{\odot} \rm{yr}^{-1} / \rm{erg} \, \rm{s}^{-1} \rm{Hz}^{-1}$] \citep{Madau-Dickinson14}. As marked, the dashed and solid lines show model results for the intrinsic and dust-attenuated values of the UV Luminosity. The different colors show results for the UV magnitude limits marked so as to be able to compare to the observations shown using points. Finally, the solid gray line shows the results from our extreme model using $f_*^{\rm{eff}}=100\%$, and a magnitude threshold of -19. The different points show observational data from \citet[][diamonds]{Donnan23} who use a magnitude threshold of -17, from \citet[][squares]{McLeod16} who use a magnitude threshold of -17.7 and from \citet[][triangles; red for fiducial, and orange, purple and olive-green for ``robust", ``solid" and ``possible" literature detections, respectively]{Bouwens23_solid}, who use a magnitude threshold of -19.}
  \label{fig:uv_density_dust}
\end{figure}

\subsection{The intrinsic and dust-attenuated UV luminosity density}
\label{sec_uvld}

We now show the redshift evolution of the UV luminosity density ($\rho_{\rm UV}$), for both the intrinsic and dust-attenuated cases, obtained for a number of different magnitude thresholds as shown in Fig. \ref{fig:uv_density_dust}. To compare to observations, we convert the UV luminosity to a star formation rate (SFR) using a conversion factor of $\kappa_{\rm UV} = \psi/L_{\rm UV} = 1.15 \times 10^{-28}$ \citep[$\rm{M}_{\odot} \rm{yr}^{-1} / \rm{erg} \, \rm{s}^{-1} \rm{Hz}^{-1}$;][]{Madau-Dickinson14}.

Integrating over all galaxies, the intrinsic UV luminosity density decreases by about three orders of magnitude from $\rho_{\rm UV} \sim 10^{26.5}$ to $10^{23.8} {\rm erg~s^{-1} Hz^{-1} cMpc^{-3}}$ between $z \sim 4.5$ and 18. 
The $\rho_{\rm UV}$ contribution of faint galaxies (with $-15<\muv<-13$) increases with redshift with values of about $12,31,69\%$ at $z \sim 7,10$ and $15$, respectively. On the other hand, the contribution of bright ($\muv \lsim -19$) galaxies to the total UV luminosity density slowly decreases with increasing redshift, being $27,7,0.4\%$ at $z \sim 7,10$ and $15$, respectively. As seen from the same figure, dust has a sensible impact only at $z\lsim 6$ for the global population and at $z \lsim 10$ for bright sources with $\muv \lsim -19$.  

Our prediction of the luminosity density is in good agreement with observed data-sets up to $z \sim 10$ integrating down to a number of magnitude thresholds ranging between $\muv \sim -17$ to $-19$ \citep[e.g.][]{Donnan23,McLeod16,Bouwens23_solid} as detailed in Fig. \ref{fig:uv_density_dust}. At $z>13$, we compare our results with available data from \citet{Bouwens23_solid}, who provide $\rho_{\rm UV}$ values for their own recent JWST detections in addition to a compilation of public JWST data that they label ``robust", ``solid" and ``possible". As seen from the same figure, the UV luminosity density from their data as well as the ``robust" data-set lie about 0.5 dex above our predicted values at $z \sim 13$, with the ``solid" and ``possible" data-sets being almost two orders of magnitude above our model values. With these data-sets effectively showing the same number density at $z \sim 13-17$, by $z \sim 17$, all of these observations lie orders of magnitude above the predicted luminosity density values. As a sanity check, we also compare these observations to our ``maximal" model (no feedback, $\feff=1$). While we find such an upper limit to be in accord with their observations as well as the ``solid" data-set, it is still lower than the value inferred from the tentative ``possible" data-set. This again leads to the conclusion that spectroscopic confirmations are crucially required to validate the redshift and nature of these highest-redshift sources. 

\begin{figure}
  \resizebox{\hsize}{!}{\includegraphics{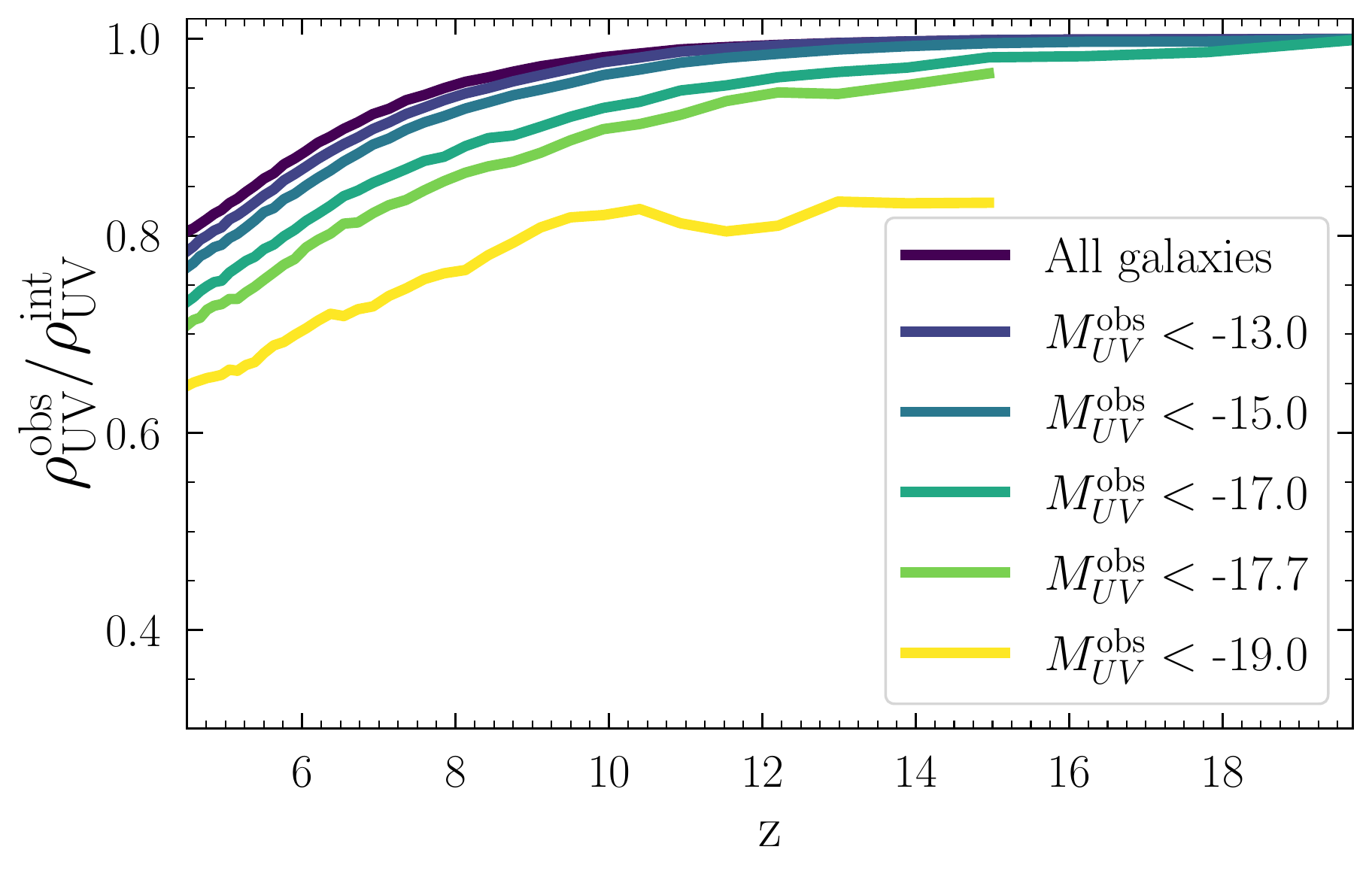}}
  \caption{As a function of redshift, for $z \sim 5-20$, we show the ratio of the dust-attenuated UV luminosity density to the total intrinsic UV Luminosity density. The different lines show results for the different dust-attenuated UV magnitudes marked.}
  \label{fig:sfr_uv_fraction}
\end{figure}

\begin{figure*}
  \resizebox{\hsize}{!}{\includegraphics{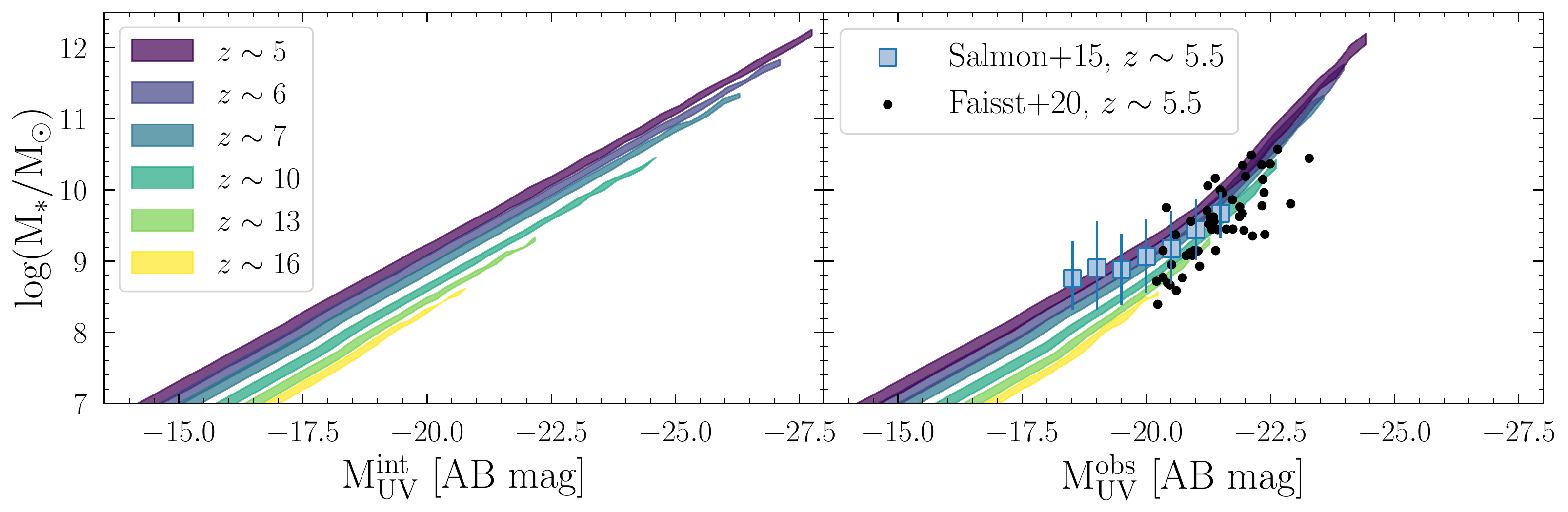}}
  \caption{The total stellar mass as a function of the intrinsic ({\it left panel}) and dust-attenuated ({\it right panel}) absolute magnitude at $z \sim 5-16$, as shown with the different shaded curves. The area of each curve represents the extent from the \sixteenth percentile to the \eightyfourth percentile. On the right panel, the blue squares represent CANDELS galaxies \citep{Grogin11,Koekemoer11} compiled in \citet{Salmon15} and the black dots are ALPINE galaxies from \citet{Faisst20}. Both data sets are taken as reported in \citet{Khusanova21}.} 
  \label{fig:stellar_mass_mag}
\end{figure*}

We quantify the effects of dust in Fig. \ref{fig:sfr_uv_fraction} where we show the population averaged fraction of UV light that manages to escape from galaxies ($\rho_{\rm UV}^{\rm obs}/\rho_{\rm UV}^{\rm int}$), for different magnitude thresholds. Accounting for all galaxies, the star-formation rate density based on the UV would miss 17\% of the actual star-formation rate density at $z\sim 5$ which decreases to 2\% at $z \gsim 10$. However, only considering bright galaxies with $\rm{M_{UV}} < -19$, UV selection would miss 34\% at $z \sim 5$, decreasing to 17\% by $z \sim 10$; this is in excellent accord with $30-60\%$ of the SFR being missed in the UV at $z \sim 7$ due to dust attenuation as inferred by ALMA REBELS results \citep{Algera2023a}. The UV photons absorbed by dust are re-emitted in the FIR, which we discuss in Sec. \ref{sec:dust_detectability}.

\subsection{Redshift evolution of the stellar mass - UV luminosity relation}
\label{sec_msuv}

We now discuss the mass-to-light relation between the intrinsic and observed UV magnitudes and the total stellar mass, at $z \sim 5-16$, as shown in Fig. \ref{fig:stellar_mass_mag}. For $M_* \gsim 10^7\msun$ galaxies, $\muv^{\rm int}$ effectively scales with the stellar mass. This is because such galaxies reside in massive halos (with $M_h \gsim 10^{9.5}\msun$) at all the redshifts considered and are therefore efficient star-formers with a fixed efficiency of $\feff = f_* = 0.15$. Moreover, at a fixed $\muv^{int}$ value, the associated stellar mass increases with decreasing redshift. This is because galaxies have lower gas fractions with decreasing redshift \citep[due to more generations of feedback-limited progenitors;][]{dayal2014} which results in lower star formation rates. The redshift-dependent relation between the intrinsic magnitude and the stellar mass is well fit by:
\be \label{eq:mint_to_mstar}
\log(\rm{M_*/M_{\odot}}) = -0.4*M_{\rm{UV}}^{\rm{int}} + 1.495 - 0.0797z. 
\ee 
From this relation, we see that $\muv^{\rm int} \sim -19$ corresponds to $M_* \sim 10^{8.7}\msun$ at $z \sim 5$ which drops by about an order of magnitude to $M_* \sim 10^{7.8}\msun$ by $z \sim 16$. We also note that the above relation implies a linear scaling between the stellar mass and the SFR.

We then discuss the $\muv^{\rm obs}-M_*$ relation shown for $z \sim 5-16$ in the right panel of the same figure. Given their low dust masses and associated dust attenuation, discussed in detail in Sec. \ref{sec:dust_properties}, the $\muv^{\rm obs}-M_*$ relation follows the intrinsic UV magnitude-stellar mass relation for $\muv^{\rm obs} \gsim -20$ at all the redshifts considered. However, as a result of the increasing dust attenuation with increasing mass, the $\muv^{\rm obs}-
M_*$ relation shows an upturn for brighter systems. For example, galaxies with $M_* \sim 10^{10}\msun$ at $z \sim 5$ show an observed magnitude of about -22.5 which is a magnitude fainter than the intrinsic magnitude. Similarly, the most massive systems, with $M_* \sim 10^{12}\msun$ show an observed magnitude ($\sim -24$) which is 2.5 magnitudes fainter than the intrinsic $\muv$ value. 
We compare our predicted stellar mass - magnitude relation with values reported by \cite{Salmon15} and \cite{Faisst20} for CANDELS \citep{Grogin11,Koekemoer11} and ALPINE galaxies, respectively, at $z \sim 5.5$. Our results match well with data from \cite{Salmon15}, which also show an upturn of the relation for brighter systems. Compared to \cite{Faisst20}, our results predict slightly larger stellar masses for a given observed magnitude on average, while still being compatible with their most massive galaxies.
The escape fraction of continuum photons is quantified in Sec. \ref{sec:uvfesc} and can be used to link the $M_*$ and $\muv^{\rm obs}$ values at any given redshift.

\subsection{The stellar mass function and stellar mass density at $z \sim 5-20$}
\label{sec_smf}
\begin{figure*}
  \resizebox{\hsize}{!}{\includegraphics{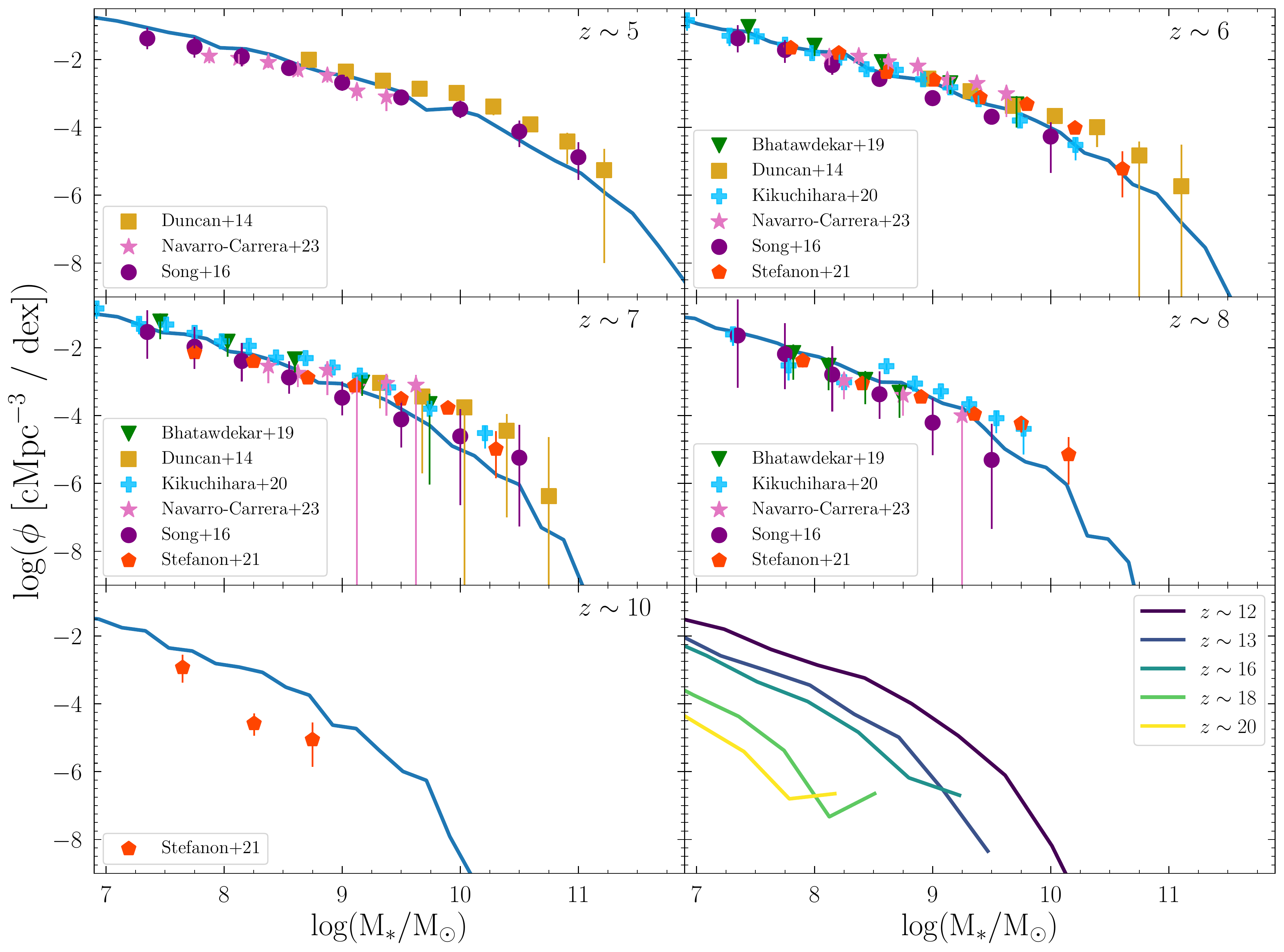}}
  \caption{The redshift evolution of the SMF predicted by our model (solid line) at $z \sim 5-20$, as marked. In each panel, we compare our model results with a compilation of observational results shown by points \citep[including][]{Duncan14, Song16, Bhatawdekar19, Kikuchihara20, Stefanon21, Navarro-Carrera23}, as marked. All of the observational data-sets have been re-normalised to a Kroupa IMF. The bottom right panel shows the model predictions at $z \sim 12-20$,  where there is no observational data yet.}
  \label{fig:smf}
\end{figure*}

\begin{figure}
  \resizebox{\hsize}{!}{\includegraphics{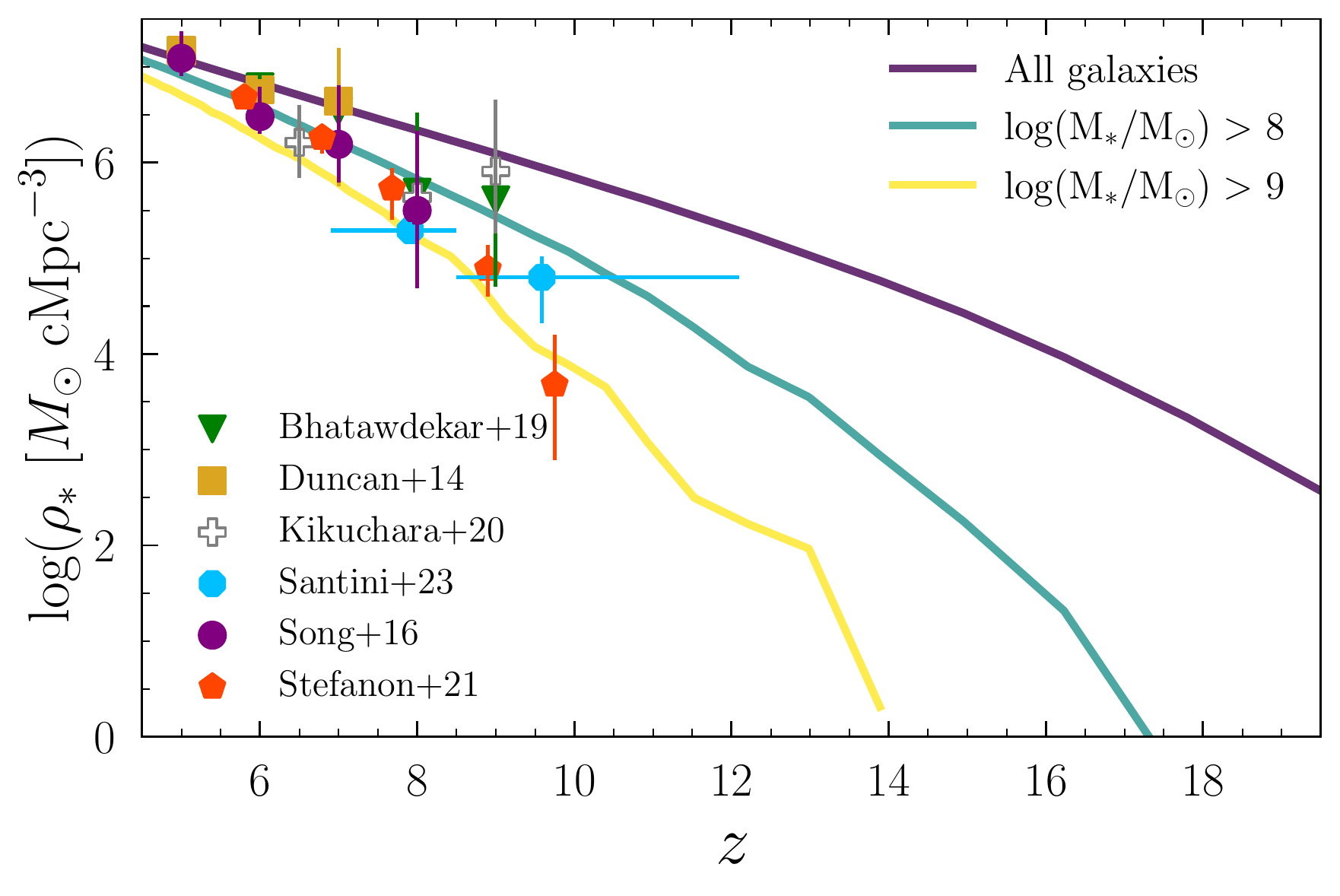}}
  \caption{The redshift evolution of the stellar mass density between $z \sim 5-19$ for different mass cuts, including all galaxies and integrating above $M_* \gsim 10^{8}$ and $10^9 M_\odot$, as marked. We compare our model results with the observationally inferred values (integrating down to a mass limit of $10^{8}M_\odot$) from \citet[][yellow squares]{Duncan14}, \citet[][purple circles]{Song16}, \citet[][green triangles]{Bhatawdekar19}, \citet[][empty crosses]{Kikuchihara20}, \citet[][red pentagons]{Stefanon21} and the recent JWST results from \citet[][blue circles]{Santini23} which have all be re-normalised to a Kroupa IMF.}
  \label{fig:smd}
\end{figure}

We now present our predictions of the stellar mass function, at $z \sim 5 - 20$, as shown in Fig. \ref{fig:smf}. We compare our results at $z \sim 5-10$ with a number of observational data-sets that are in good agreement within error bars \citep[from][]{Duncan14,Song16,Bhatawdekar19,Kikuchihara20, Stefanon21, Navarro-Carrera23} when re-normalised to a Kroupa IMF. By construction, the theoretical SMF is a good match to the data for $M_* \sim 10^{7-11.25}\msun$ at $z \sim 5-8$. Again, while the low-mass end is mostly determined by SNII feedback and the $f_w$ parameter, the high-mass end is determined by the star formation efficiency ($\feff = f_*$) in these massive sources. The transition between the two regimes takes place at $\rm{M_*} = 10^8 (10^{7.6}) M_{\odot}$ corresponding to $\rm{M_{UV}^{obs}} = -17.7 (-18.4)$ at redshift 5 (15).
At $z \sim 10$, however, the theoretical SMF over-predicts the number density for $M_* \sim 10^{8.25-8.75}\msun$ sources from \citet{Stefanon21} by as much as an order of magnitude. This could be due to a number of reasons including low number statistics, the low number of photometric data points, or the assumption of a constant star formation history leading to an under-estimation of the stellar mass observationally \citep{Topping22}. In the same figure, we also show the SMF predicted by our model at $z \sim 12-20$. As seen, both the amplitude and the mass range of the SMF decrease with increasing redshift. For example, for $M_* \sim 10^7 \msun$, the number density falls by about 3.25 orders of magnitude between $z \sim 12$ and $20$, from a value of $10^{-2}$ to $10^{-5.25} {\rm cMpc^{-3} dex^{-1}}$. Also, considering down to number densities of $10^{-8.5} cMpc^{-3}$, while the SMF extends between $10^{7-9.5}\msun$ at $z \sim 12$, this narrows to a maximum mass of $10^{7.5}\msun$ by $z \sim 20$.   

We then show the associated stellar mass density (SMD, $\rho_*$) in Fig. \ref{fig:smd}. Starting with all galaxies, the SMD decreases by about 4.5 orders of magnitude from $\rho_* \sim 10^{7.2}\msun {\rm cMpc^{-3}}$ at $z\sim 4.5$ to $\rho_* \sim 10^{2.5}\msun {\rm cMpc^{-3}}$ by $z \sim 19.5$. We also show the SMD integrating down to mass limits of $10^8$ and $10^9\msun$; the former mass limit corresponds to the limiting mass for most observational studies \citep{Duncan14,Bhatawdekar19,Kikuchihara20, Stefanon21,Santini23}. Galaxies more massive than $M_* = (10^8)~10^9\msun$ contribute roughly 74\%~(49\%) to the total SMD at $z \sim 4.5$. The number density of $M_* = (10^8)~10^9\msun$ sources fall off with redshift such that they contribute less than $4\%~(0.1\%)$ to the SMD by $z \sim 12$. 

It is interesting to see that within error bars, all of the SMD data points are in good agreement with each other despite their varying assumptions regarding the star formation histories, metallicities and selection techniques at $z <10$. At $z \sim 10$, given the paucity of data, the observed SMD values show a dispersion of about an order of magnitude, ranging between $10^{3.5-4.5}\msun {\rm cMpc^{-3}}$. As might be expected from the SMF discussed above, the theoretical SMD values are in good agreement with the observations within error bars at $z < 10$. At $z \sim 10$, however the observed SMD values are about 0.5 to 1.0 dex lower than the predictions of the theoretical model for galaxies more massive than $M_* = 10^8\msun$. As discussed above, this could be due to incompleteness in the observed sample as well as an under-estimation of the stellar mass due to the assumption of a constant star formation history. One can also not rule out the fact that the IMF might evolve and become increasing top heavy with increasing redshift \citep[see e.g][]{pacucci2022,chon2021}, yielding a lower mass-to-light ratio as compared to our model. JWST confirmations of high-redshift sources will be crucial in constraining the SMD and baselining and validating theoretical models at these high-redshifts, although such mass determinations might not extend to very faint sources.

\section{Dust properties in the first billion years} \label{sec:dust_properties}

In this section we study the dust properties of early galaxies including the dust-stellar mass relation (Sec. \ref{sec_mdms}), the escape fraction of UV photons unattenuated by dust (Sec. \ref{sec:uvfesc}) and the associated dust temperatures (Sec. \ref{sec_td}).

\begin{figure}
  \resizebox{\hsize}{!}{\includegraphics{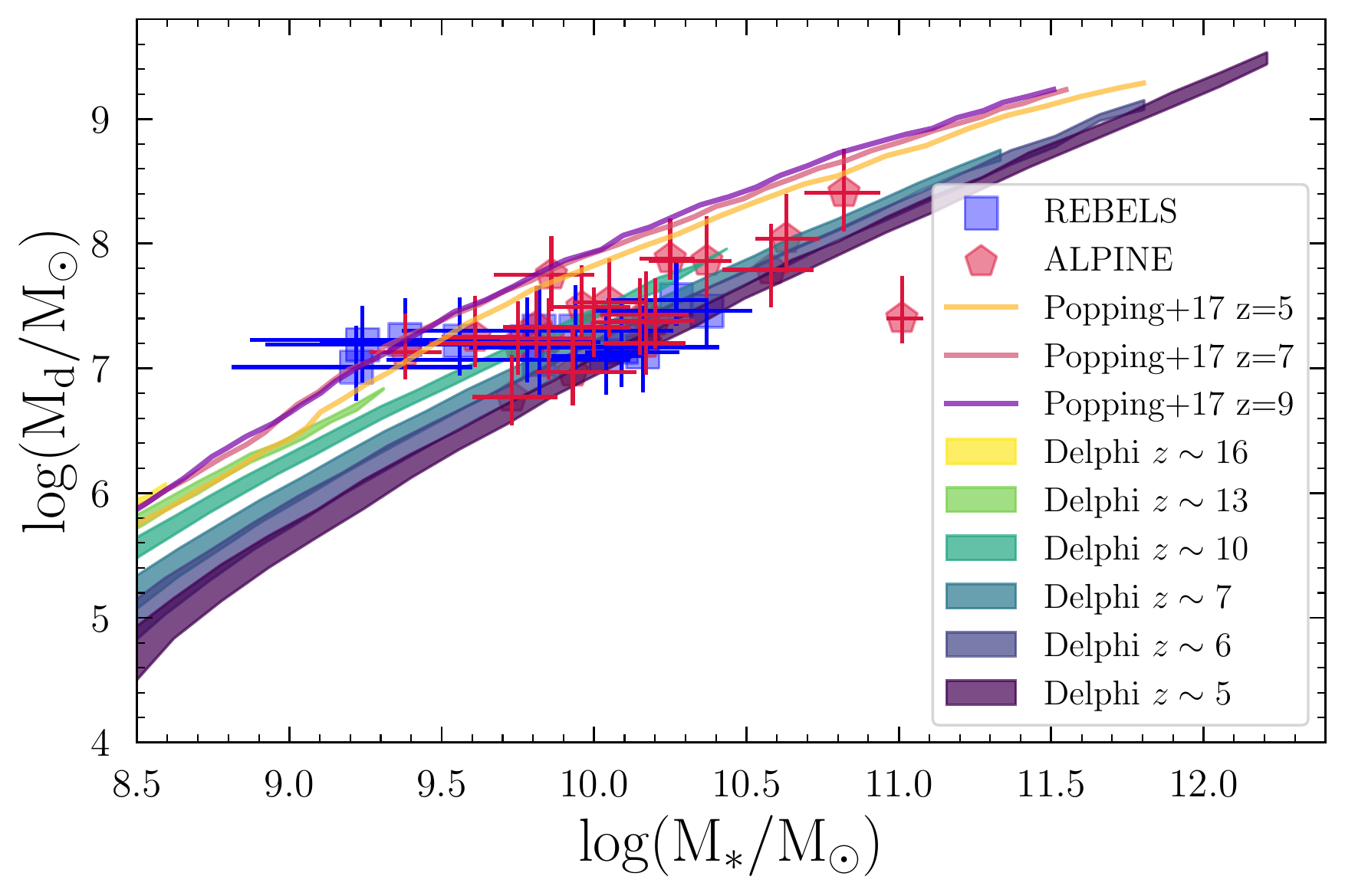}}
  \caption{The dust mass as a function of the total stellar mass for $z \sim 5-16$, as marked; the area of each curve demarcates the extent from the \sixteenth percentile to the \eightyfourth percentile. The yellow, orange and purple lines show the median results from the fiducial model of \citet{Popping17} at $z \sim$5, 7 and 9. Finally, as marked, the different points show results from the ALMA REBELS \citep[blue squares;][]{REBELS} and ALPINE surveys \citep[pink pentagons;][]{Fudamoto20}.}
  \label{fig:dust_mass_stellar_mass}
\end{figure}

\subsection{The relation between dust and stellar mass in the first billion years}
\label{sec_mdms}
We start by showing the relation between dust mass ($M_d$) and stellar mass in Fig. \ref{fig:dust_mass_stellar_mass}. Firstly, we find a linear relation linking $M_*$ and $M_d$ at all $z \sim 5-16$ \citep[see also Sec. 3,][]{dayal2022}. This is driven by the fact that the SNII dust production rate is proportional to the SFR that scales with $M_*$ in our model. Further, all of the processes of astration, destruction and ejection also scale with the SFR \citep[see Sec. 2.1][]{dayal2022} with ISM grain growth on a 30 Myr timescale only resulting in a small contribution to the total dust mass. Secondly, at a fixed stellar mass galaxies show a dust mass that increases with increasing redshift. This can be explained by the fact that the halo rotational velocity ($v_c$) increases with increasing redshift for a given halo mass. This leads to an increase in $\fej$ which leads to a higher star-formation rate for feedback-limited galaxies (which form stars at an efficiency of $\fej$) resulting in a higher dust mass. Furthermore, by combining Equations \ref{eq:mstar}, \ref{eq:def_fej} and \ref{eq:Mej}, we see that an increased $v_c$ leads to a decrease in the ejected gas mass, for both efficient star formers and feedback limited galaxies, resulting in both retaining a larger fraction of their gas and dust content. Indeed, for $M_* \gsim 10^{8.5}\msun$, we find a linear relation linking $M_d$ and $M_*$ at all $z \sim 5-16$ such that
\be \label{eq:mstar_to_mdust}
\log(M_d) = 1.194\log(M_*) + 0.0975z - 5.433.
\ee
For galaxies of $M_* \sim 10^{9.5} \msun$, our model results in a dust mass of about $10^{6.4} ~ (10^{6.9})\msun$ and a dust-to-stellar mass ratio of about $0.08\% (0.25\%)$ at $z \sim 5 ~(10)$. This increases to a dust mass of about $10^{8.2}~ (10^{8.7}) \msun$ and a dust-to-stellar mass ratio of about $0.16\% (0.5\%)$ at $z \sim 5 ~(10)$ for massive galaxies with $M_* \sim 10^{11}\msun$.

We then compare our results to the dust masses inferred for $M_*\sim 10^{9-11}\msun$ galaxies at $z \sim 5$ and $7$ from the ALMA ALPINE \citep{Fudamoto20} and REBELS \citep{REBELS} surveys, respectively. We note two key caveats involved in these observational data-sets: firstly, given most of these sources are detected in a single ALMA band, a dust temperature has to be assumed in order to obtain a dust mass \citep[see discussion in][]{Sommovigo22a}. Further, the star formation history used can significantly affect the inferred stellar masses \citep[see e.g.][]{Topping22}. Despite these caveats, within error bars our model results at $z \sim 5$ are in good accord with the ALPINE results, except perhaps for two galaxies, the lowest-mass and highest-mass sources. Further, the REBELS sample finds a rather flat distribution of the dust masses as a function of the stellar mass \citep[see Sec. 3,][]{dayal2022} as compared to the linear relation found by the theoretical model. Possible solutions could lie in the stellar masses being under-estimated using the assumption of a constant star formation history \citep{Topping22} or higher dust temperatures that could push down the associated dust masses. We also note that ALPINE sources (at lower redshifts) seem to indicate slightly larger dust masses for a given stellar mass as compared to REBELS galaxies. Although contrary to the trends we find, we urge caution in light of the low numbers of sources and caveats on the dust temperatures in addition to the stellar mass caveats listed above. Finally, we also show results from the fiducial model of \cite{Popping17} at $z \sim$5, 7 and 9. As shown, they find a dust mass than is larger than ours by a factor of about 6. This is due to the smaller dust growth timescale that they use, resulting in a dust mass dominated by dust growth, while dust growth has a sub-dominant impact in our models, as shown in \cite{dayal2022}. Similar to us, \cite{Popping17} find that at a fixed stellar mass, dust masses are slightly larger at higher redshifts, as shown in Fig. \ref{fig:dust_mass_stellar_mass}.

\begin{figure*}
  \resizebox{\hsize}{!}{\includegraphics{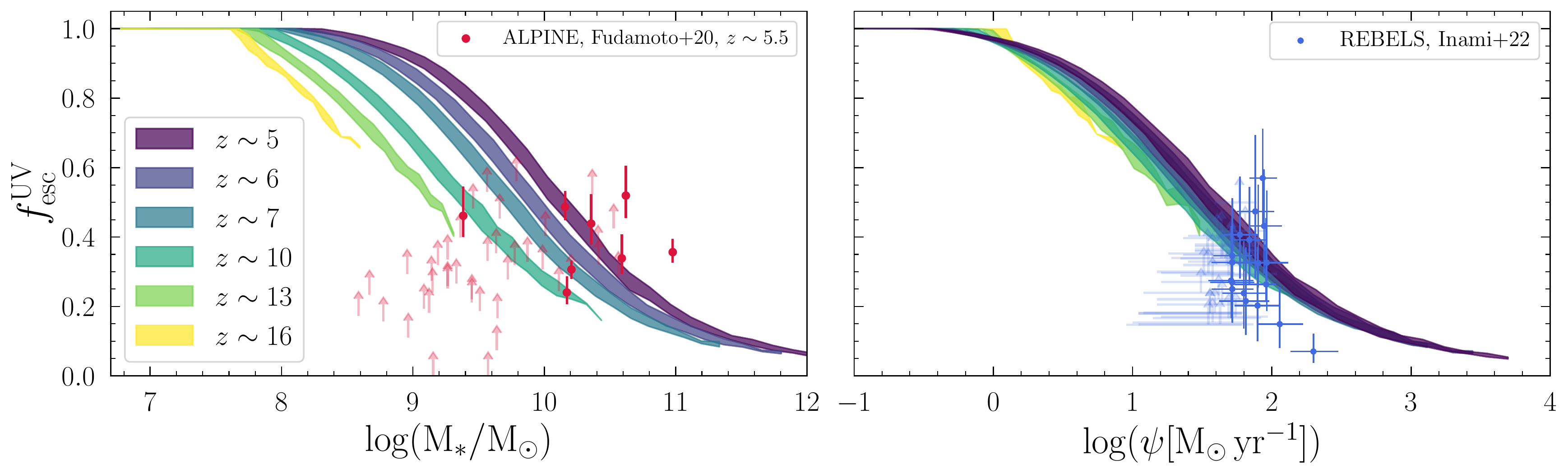}}
  \caption{Model results showing the escape fraction of non-ionizing UV photons ($\fescuv$) as a function of the total stellar mass ({\it left panel}) and the star-formation rate ({\it right panel}). The curves 
  show the results for $z \sim 5-16$, as marked and demarcate the extent from the \sixteenth percentile to the \eightyfourth percentile. Overplotted are observational results from ALPINE galaxies \citep{Fudamoto20} and REBELS galaxies \citep{Inami22}, on the left and right panel, respectively. Transparent arrows indicate lower limits.}
  \label{fig:uv_fesc_stellar_mass_sfr}
\end{figure*}

\subsection{The evolution of the UV escape fraction}
\label{sec:uvfesc}
We now look at the relation between the fraction of UV photons that can escape a galaxy unattenuated by dust ($\fescuv$) and the stellar mass and SFR. The UV escape fraction can also be interpreted as the ratio of the SFR observed in the UV ($\psi_{\rm UV}$) to the total intrinsic SFR ($\psi$) i.e. $\fescuv = \psi_{\rm UV}/\psi$. We start by studying $\fescuv$ as a function of the stellar mass in (the left panel of) Fig. \ref{fig:uv_fesc_stellar_mass_sfr}. We find two key trends: at a given redshift, $\fescuv$ decreases with increasing $M_*$ given the increasing dust masses of more massive systems. For example, at $z \sim 5$, $\fescuv$ decreases from $\sim 1$ for $M_* \lsim 10^{8.5}\msun$ to $\sim 0.05$ for $M_* \sim 10^{12}\msun$ systems. As we go to higher redshifts, the stellar mass range naturally narrows: for example, by $z \sim 16$, the most massive systems only have $M_* \sim 10^{8.5}\msun$ and $\fescuv \sim 0.65$. Secondly, as noted in the previous section, for a given stellar mass, the dust mass increases slightly with increasing redshift. Further, galaxies of a given stellar mass are hosted in slightly lower-mass halos (i.e with smaller virial radii) with increasing redshift. This, coupled with our assumptions of the gas and dust radius being effectively constant with redshift for a fixed halo mass result in a decrease in $\fescuv$ with increasing redshift for a given stellar mass. Indeed, considering $M_* \sim 10^{9.5}\msun$, $\fescuv$ decreases from $\sim 0.8$ at $z \sim 5$ to $\sim 0.5$ by $z \sim 10$. 
Our redshift-dependent relation between $\fescuv$ and $M_*$ at $z \sim 5-16$ is quantified as:
\be \label{eq:mstar_to_fesc}
\fescuv = \frac{1}{2} \Bigl(1 - \tanh{\bigl[\alpha(z) (\log(M_*)-M_0(z))\bigr]} \Bigr),
\ee
where $\alpha(z) = 0.931 + \exp{\bigl(0.447z-7.842\bigr)}$ and $M_0(z) = 10.739 - 0.124z$. We compare our results with ALPINE galaxies from \cite{Fudamoto20}, and find a relatively good match, although observations show a significantly larger scatter compared to the theoretical results.

We then also show $\fescuv$ as a function of the total SFR in the (right panel of the) same figure. Interestingly, the $\fescuv-\psi$ relation does not show any significant evolution with redshift, over $z \sim 5-16$. This is driven by the fact that both the dust mass and SFR scale with the stellar mass in the same way as a function of redshift. I.e. for a given stellar mass, the increase in dust mass with increasing redshift is matched by an increase in the SFR resulting in a roughly constant $\fescuv-\psi$ relation. We find $\fescuv\sim 1$ for $\psi \lsim 1 \msun {\rm yr^{-1}}$ which decreases to $\sim 0.1$ for $\psi \sim 1000 \msun {\rm yr^{-1}}$. At $z \sim 7$, we find $\fescuv \sim 0.53-0.17$ for $\psi \sim 30-300 \msun {\rm yr^{-1}}$ i.e. about $47-83\%$ of the UV luminosity of such sources is suppressed due to dust attenuation. 
The $\fescuv-\psi$ relation can be quantified as:
\be \label{eq:sfr_to_fesc}
\fescuv = \frac{1}{2} \Bigl(1 - \tanh{\bigl[A(z) (\log(\psi)-\psi_0(z))\bigr]} \Bigr),
\ee
where $A(z) = 0.899 + \exp{\bigl(0.411z-7.189\bigr)}$ and $\psi_0(z) = 1.903 - 0.052z$. We compare our $\fescuv-\psi$ relation with results from the ALPINE REBELS survey presented in \cite{Inami22}. The observations are again more scattered than our model with a number of highly star forming galaxies lying below the predicted $\fescuv$ value \citep[which could be driven by e.g. dust being more clumped around star forming regions as also discussed in][]{dayal2022}. Overall, however, the theoretical results are in reasonable agreement with these observations.

Finally, we also derive a relation between the observed UV magnitude and $\fescuv$ such that
\be \label{eq:mobs_to_fesc}
\fescuv = \frac{1}{2} \Bigl(1 - \tanh{\bigl[\xi(z) (\rm{M_{UV}^{obs}}-\rm{M_{UV,0}}(z))\bigr]} \Bigr),
\ee
where $\xi(z) = -6.23 \cdot 10^{-1} + 1.58 \cdot 10^{-3} z^2 - 7.52 \cdot 10^{-6} z^4$ and $\rm{M_{UV,0}}(z) = -22.103 + 0.09z$. This is to allow a conversion between $\fescuv$ and galaxy properties including the intrinsic UV magnitude and dust masses using Eqns. \ref{eq:mint_to_mstar} - \ref{eq:mstar_to_mdust}. 

\subsection{The redshift evolution of dust temperatures}
\label{sec_td}
\begin{figure}
  \resizebox{\hsize}{!}{\includegraphics{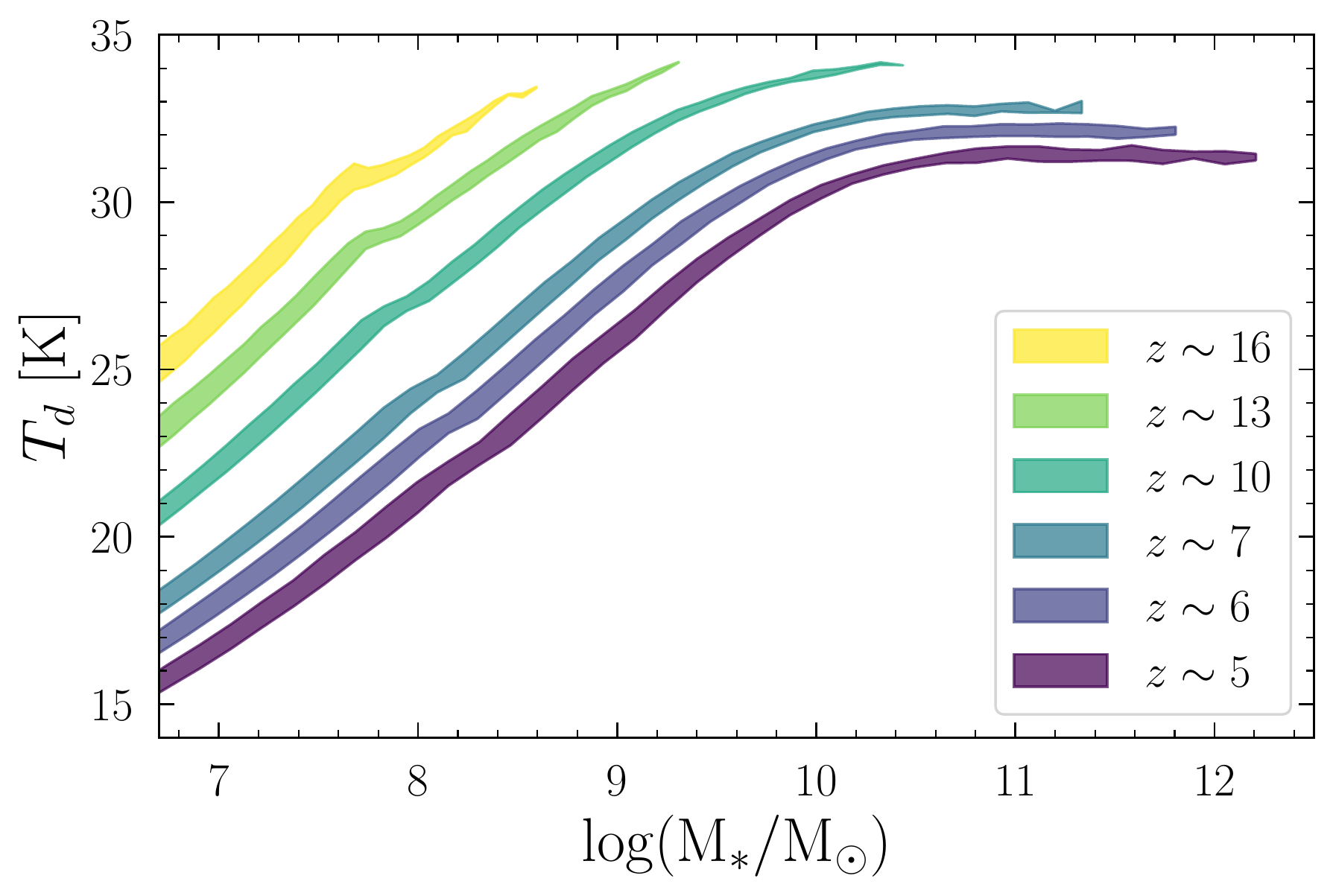}}
  \caption{The dust temperature ($T_d$) as a function of the stellar mass for $z \sim 5-16$, as marked; the area of each curve represents the extent from the \sixteenth percentile to the \eightyfourth percentile of the distribution. See text in Sec. \ref{sec:UV_IR} for details on the calculation of the dust temperature.  }
  \label{fig:dust_T_stellar_mass}
\end{figure}

Next, we study the dust temperature ($T_d$), which is a measure of how intensely the dust is heated by UV radiation. We show the dust temperature as a function of stellar mass for $z \sim 5-16$ in Fig. \ref{fig:dust_T_stellar_mass}. As seen, we find that $T_d$ increases with an increase in $M_*$. This is to be expected given that while the intrinsic UV luminosity scales with $M_*$, more massive galaxies also show lower $\fescuv$ values, allowing for more heating of the dust mass (as seen in Fig. \ref{fig:uv_fesc_stellar_mass_sfr}). For example, at $z \sim 5$, $T_d$ increases from $\sim 17$ to $31{\rm K}$ as $M_*$ increases from $10^7$ to $10^{10}\msun$. However, $T_d$ saturates for $M_* \gsim 10^{10}\msun$ - this is because of the saturation in $\fescuv$ seen above in Sec. \ref{sec:uvfesc}. We also find that at a fixed $M_*$ value, $T_d$ increases with increasing redshift. This is driven by the fact that galaxies of a given stellar mass have both a higher SFR and a smaller $\fescuv$ value with increasing redshift. This results in a larger fraction of the UV photons being absorbed from intrinsically brighter galaxies, resulting in both higher FIR luminosities and dust temperatures.

For galaxies with $M_* \sim 10^{9.5-11}\msun$, we calculate values of $T_d \sim 27-34{\rm K}$ at $z \sim 5-7$. These are lower than the average values of $T_d \sim 48\pm 8$ derived for the ALMA ALPINE sample \citep{Sommovigo22a}, and the values of $T_d \sim 47\pm 6 - 52\pm11 {\rm K}$ derived for REBELS sources \citep{Sommovigo22b, Ferrara22}, respectively. However, multi-band ALMA observations of three massive galaxies, with $M_* \sim 10^{10}\msun$ in the REBELS survey hint at lower dust temperatures of $T_d \sim 30-35{\rm K}$ \citep{Algera23}. These are in perfect agreement with the average value of about 33K we predict for such sources. An outstanding issue, however, is that such low dust temperatures result in higher dust masses that are more compatible with an unphysical ``maximal" dust model where each SNII is required to produce $1 \msun$ of dust, dust is required to grow in the ISM on a timescale of 0.3 Myr and dust can neither be destroyed nor ejected \citep[see Sec. 2.1][]{dayal2022}.

The need of the hour is multi-band ALMA detections of such high-redshift sources to get better constraints on their dust temperatures. In addition, our simplistic model misses a number of crucial effects such as the fact that dust is probably clumped in the ISM - indeed, concentrated clumps of dust around star-forming regions would have higher dust temperatures than the fully diffuse dust component calculated here. Furthermore, we caution that we have shown the intrinsic dust temperatures that do not account for the CMB temperature. As noted before, the CMB creates a temperature floor for both dust and gas temperatures \citep[e.g.][]{dacunha2013} that scales as $T(z) = 2.732 (1+z)$ K in addition to creating the background against which dust emission is observed. This corresponds to $T(z) \sim 30, 38.2$ and $46.4$ K at $z = 10, 13$ and $16$, respectively. As seen from Fig. \ref{fig:dust_T_stellar_mass} above, as per our calculations, this would imply that only galaxies with $M_* \gsim 10^{10}\msun$ would be visible in terms of their dust emission at $z \sim 10$ with no galaxies visible in dust emission at $z \gsim 13$.

\section{The theoretical FIR LF at redshifts 5 to 20} 
\label{sec:dust_detectability}

\begin{figure}
  \resizebox{\hsize}{!}{\includegraphics{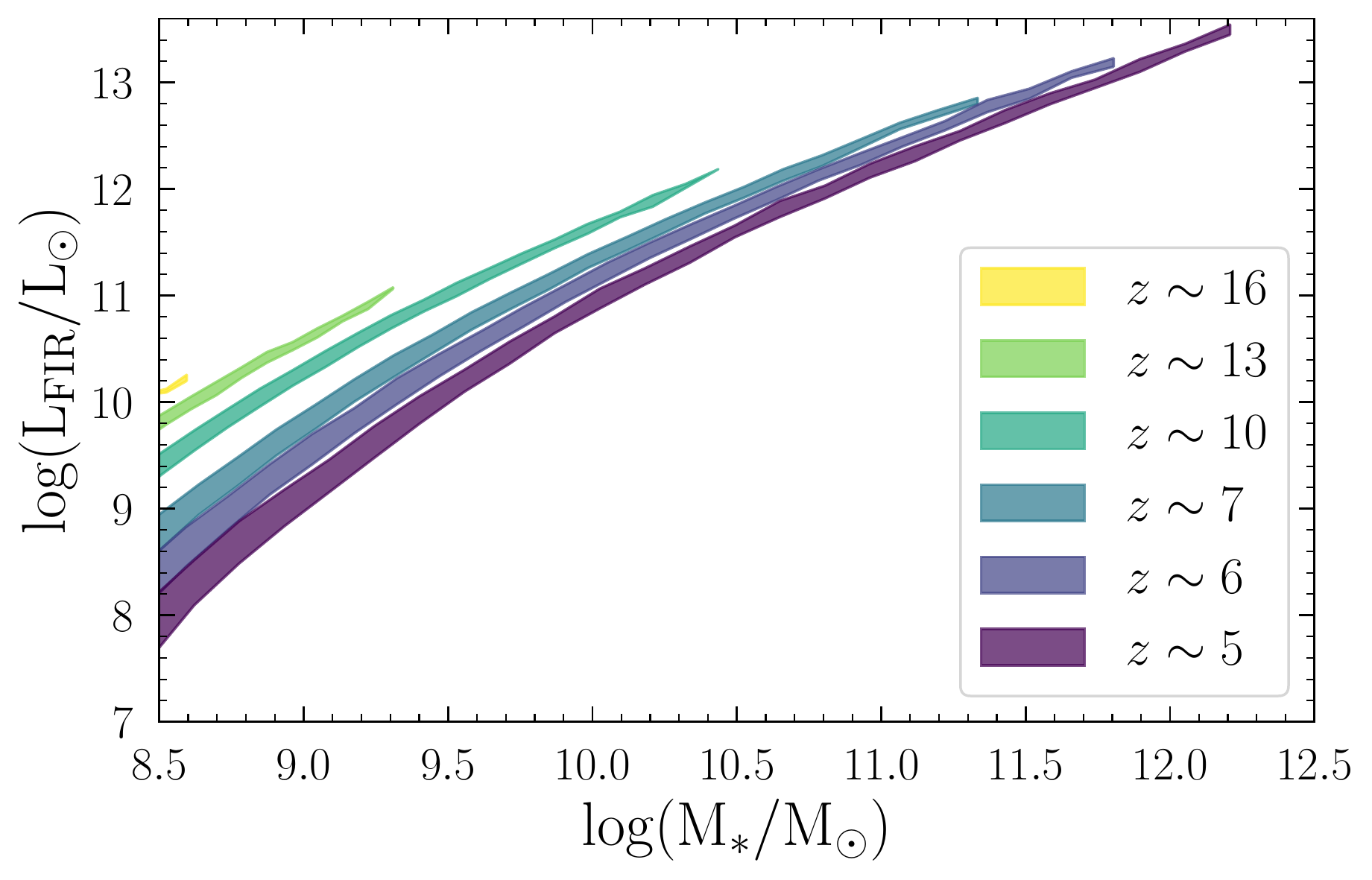}}
  \caption{Model results showing the FIR dust emission ($L_{\rm FIR}$) as a function of the stellar mass for $z \sim 5-16$, as marked; the area of each curve represents the extent from the \sixteenth percentile to the \eightyfourth percentile of the distribution. See text in Sec. \ref{sec:UV_IR} for details on the calculation of $L_{\rm FIR}$.}
  \label{fig:fir_stellar_mass}
\end{figure}

Now that we have established that the model reproduces observables in the UV and have studied the dust enrichment and attenuation of early sources, we can study their dust emission. We start by discussing the relation between the FIR luminosity ($L_{\rm{FIR}}$) and stellar mass as shown in Fig. \ref{fig:fir_stellar_mass}. We see that $L_{\rm{FIR}}$ increases with stellar mass due to the higher star-formation rate of more massive galaxies and their lower $\fescuv$ values that lead to more UV photons being absorbed by dust and re-emitted in the infrared. For galaxies with $M_* \sim 10^{9.5-11}\msun$, our model yields $L_{\rm{FIR}} \sim 10^{10.1-12.2} ~(10^{10.6-12.5})\lsun$ at $z \sim 5~(7)$. Further, as might be expected from the discussions above, for a given $M_*$ value, $L_{\rm{FIR}}$ increases with increasing redshift as a result of their larger dust masses and lower $\fescuv$ values. Indeed, by $z \sim 10$, galaxies with $M_* \sim 10^{9.5}\msun$ show FIR luminosity values as high as $L_{\rm{FIR}}\sim 10^{11}\lsun$. 

\begin{figure*}
  \resizebox{\hsize}{!}{\includegraphics{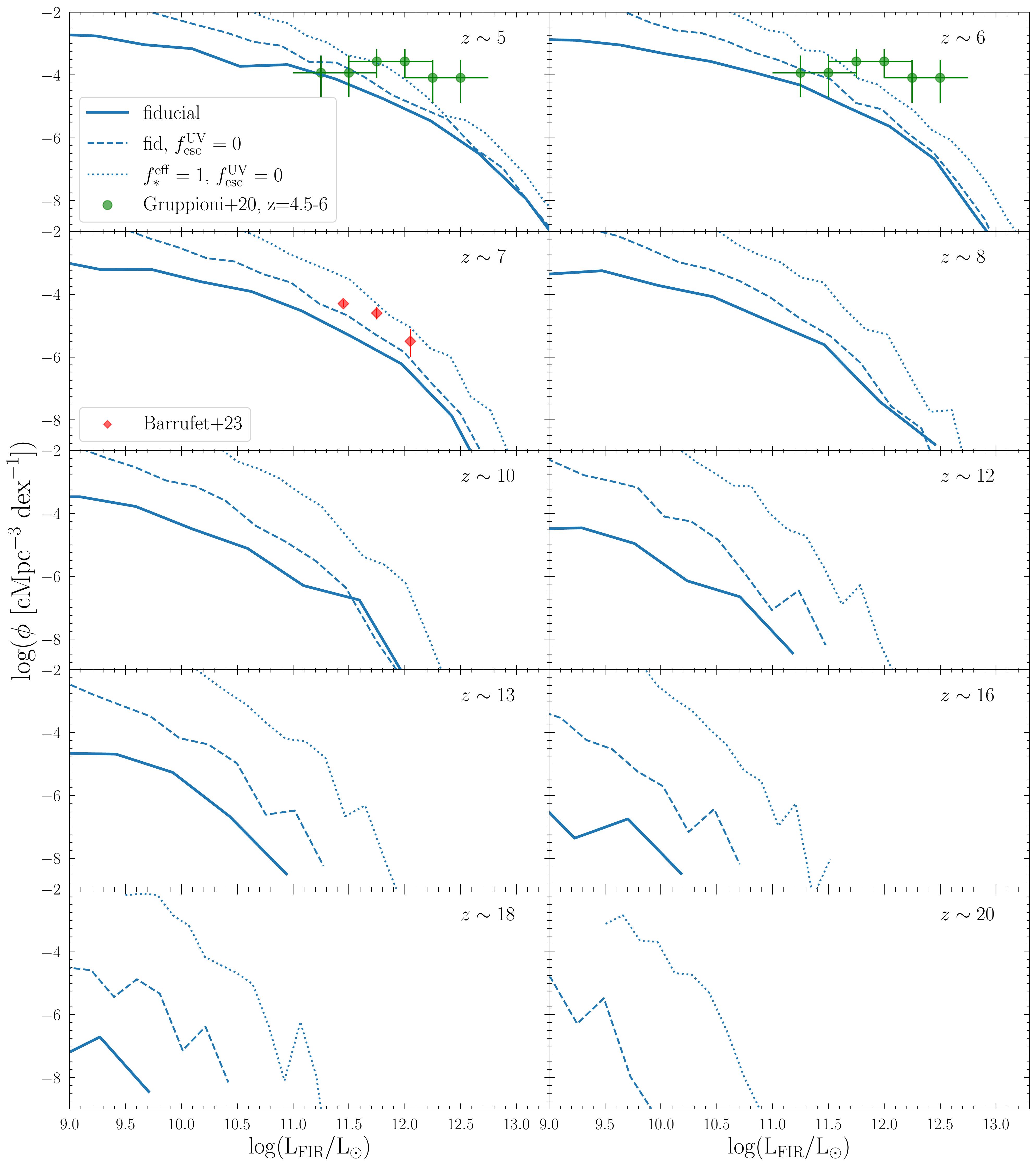}}
  \caption{The redshift evolution of the infrared LF at $z \sim 5-20$, as marked in the panels. In each panel, the solid line represents the {\it fiducial} model results, the dashed line shows results from the fiducial model assuming all UV luminosity is re-emitted in the FIR (i.e. $\fescuv=0$) and the dotted line shows the ``maximal" model with $\feff=1$ and $\fescuv=0$. We compare our model results to observed FIR LFs at $z \sim 5-6$ \citep[][green dots]{Gruppioni20} and at $z \sim 7$ \citep[][red diamonds]{Barrufet23}.}
  \label{fig:fir_lum}
\end{figure*}

We then show our resulting FIR LF at $z \sim 5-20$ in Fig. \ref{fig:fir_lum}. As might be expected, both the normalisation and the luminosity range of the FIR LF decrease with increasing redshift. For example, the number density of sources with $L_{\rm FIR} \sim 10^{9.5}\lsun$ falls from $10^{-3} {\rm cMpc^{-3}~dex^{-1}}$ at $z \sim 5$ to $10^{-7.5}{\rm cMpc^{-3}~dex^{-1}}$ by $z \sim 18$. This is because the number density associated with a given stellar mass drops-off with increasing redshift faster than the increase in FIR luminosity. Further, at $z \sim 5$, the FIR LF extends between $10^{9-13.5}\lsun$ which decreases to $10^{9-12.5}\lsun$ by $z \sim 7$ and $<10^{9.7}\lsun$ by $z\sim 18$; there is effectively no FIR LF at redshifts as high as $z \sim 20$. Indeed, at $z \sim 13$, the fiducial model yields slightly more than one per cGpc$^3$ for $L_{\rm FIR} \sim 10^{11}\lsun$ and about 10 galaxies per cGpc$^3$ of $L_{\rm FIR} \sim 10^{10}\lsun$ at $z \sim 16$. Getting significant number statistics at these early epochs therefore poses a severe challenge in the volumes that must be surveyed. 

We then compare our results to the FIR LFs inferred using ALMA data: this includes results at $z \sim 4.5-6$ from the ALPINE survey \citep{Gruppioni20} and from the REBELS survey at $z \sim 7$ \citep{Barrufet23}. We start by noting that both these samples are based on low number statistics. Further, the $z \sim 4.5-6$ data shows a number of puzzling aspects such as the flatness of the FIR LF over the observed range of $L_{\rm FIR} \sim 10^{11.25-12.5}\lsun$ and the volume density of dusty sources seems to show very little evolution at $z>2.5–3$. This could arise from a number of reasons \citep[see discussion in][]{Gruppioni20} including: (i) photometric redshift uncertainties that can induce Poissonian errors in the LF (i.e. an uncertainty in the number of objects in each bin); (ii) the sources probed might be part of an over-density; considering them as unbiased blindly detected sources would thereby lead to an overestimation of the LF; (iii) the fact that most of these sources are detected in a single ALMA band which results in uncertainties in converting such monochromatic fluxes to total FIR luminosities. The same issues also hold true for the FIR LF at $z \sim 7$. 

With these caveats in mind, we find that while the {\it fiducial} theoretical FIR LF is in good agreement with the $z \sim 4.5-6$ data in the faintest luminosity bins ($L_{\rm FIR} \sim 10^{11.25-11.5}\lsun$), it under-predicts the number density for brighter sources. The same situation arises when comparing to the data at $z \sim 7$ where our {\it fiducial} results lie below the observations by as much as an order of magnitude for the faintest sources. We therefore carry out a number of limiting calculations at all $z \sim 5-20$: (i) in the first case, we use the {\it fiducial} model for the UV luminosity but assume $\fescuv=0$ i.e. all of the UV photons are converted into FIR luminosity; (ii) in the ``maximal" case, we assume a SFE of a 100\% i.e. $\feff=1$ and $\fescuv=0$ - this yields the upper limit to the FIR LF at any redshift. 

We find that, within error bars, the observed FIR LF at $z \sim 4.5-6$ is in accord with the ``maximal" model for $L_{\rm FIR} \sim 10^{11.75-12.25}\lsun$. Puzzlingly, however, the brightest observed data point lies above this maximal model. The situation is similar at $z \sim 7$ where the observationally-inferred FIR LF is more compatible with the maximal model, at least for the brightest sources. It is therefore crucial to have spectroscopic confirmation for the redshift of these sources, and preferably multi-band ALMA observations to robustly pin-down the FIR LF at high-redshifts before we invoke unphysical extrema in galaxy formation models. 

\section{Conclusion and discussions}\label{sec:discussion}
In this work we track the dust enrichment of galaxies at $z \sim 5-20$ using the \code{DELPHI} semi-analytic model for galaxy formation. A key strength of this model is that it only invokes two mass- and redshift-independent free parameters to match of observables at $z \sim 5-9$ including the UV LF and SMF: these are the upper limit to the star-formation efficiency parameter ($f_* = 0.15$) and the fraction of SN feedback coupling to winds ($f_w = 0.06$). This model is also baselined against dust mass estimates of early galaxies from recent ALMA observations at $z \sim 5-7$. This model is used to study the impact of dust on global galaxy properties up to $z \sim 20$ - including the UV LF, SMF, UV luminosity (SFR) density. Additionally, we study the dust properties of early galaxies including the dust-to-stellar mass relation, the escape fraction of UV photons unattenuated by dust and dust temperatures before we make predictions for the dust visibility (through the FIR LF). Our key results are summarized as follows: 

\begin{itemize}

\item By construction, our model matches the observed UVLF at $z \sim 5-9$. While SNII feedback effectively shapes the faint-end of the UV LF ($\muv \gsim -20$), dust plays a key role in determining the bright-end ($\muv \lsim -21$) at $z \sim 5-10$. Further, we find that dust has no sensible impact on visibility of early galaxies at $z \gsim 12$.

\item At $z \sim 12-18$, the model significantly under-predicts both the observed UV LF and $\rho_{\rm UV}$ \citep[when comparing to e.g.][]{Donnan23,Bouwens23_solid,Harikane23}. Indeed, even a ``maximal" model with no feedback and a 100\% star formation efficiency does not produce the bright ($\muv \lsim -20$) galaxies observed at $z \sim 16-18$ \citep{Harikane23,Bouwens23_solid} and under-predicts the observationally-inferred UV luminosity density. This necessitates spectroscopic confirmations of such early sources although plausible solutions might also lie in such systems being extreme star-formers or having a top-heavy IMF.

\item While the model matches the observed SMF and SMD up to $z \sim 8$, it lies approximately 1 dex above the observed SMF (and hence the SMD) at $z \sim 10$ \citep{Stefanon21}. This might be attributed either to an incompleteness in the observational data-set or an under-estimation of the stellar masses because of an assumption of a constant star-formation history.

\item Given that SNII are the key dust factories, in our model the dust mass evolves linearly with the stellar mass at $z \sim 5-16$ such that $\log(M_d) = 1.194\log(M_*) + 0.0975z - 5.433$. As seen, for a given stellar mass, the dust mass shows an increase with redshift. This is due to the increasing star-formation rates and decreasing gas mass ejected for a fixed stellar mass with increasing redshift. 

\item The UV escape fraction $\fescuv$ decreases with stellar mass (or the star formation rate) due to the more dusty nature of massive galaxies. For example, at $z \sim 5$, $\fescuv$ decreases from $\sim 1$ for $M_* \lsim 10^{8.5}\msun$ to $\sim 0.05$ for $M_* \sim 10^{12}\msun$ systems. We also find that given their larger dust masses, galaxies of a given stellar mass show decreasing $\fescuv$ values with increasing redshift. 

\item We find that accounting for all galaxies, the star-formation rate density based on the UV would miss 17\% of the actual star-formation rate density at $z\sim 5$ which decreases to 2\% at $z \gsim 10$. However, only considering bright galaxies with $\rm{M_{UV}} < -19$, UV selection would miss 34\% at $z \sim 5$ decreasing to 17\% by $z \sim 10$; this is in excellent accord with $30-60\%$ of the SFR being missed in the UV at $z \sim 7$ due to dust attenuation as inferred by ALMA REBELS results \citep{Algera2023a}. 

\item Assuming equilibrium between the non-ionizing photons absorbed and re-radiated by dust, we find dust temperatures that increase both with stellar mass and increasing redshifts. At $z \sim 7$, we find an average temperature of 33K for galaxies with a stellar mass above $10^{10} \msun$, in good agreement with recent multi-band ALMA measurements \cite{Algera23}.

\item Finally, we predict the FIR LF at $z \sim 5-20$ and find our predictions to match to the FIR LF inferred from ALMA ALPINE results at $z\sim 5-6$ for $L_{\rm{FIR}} \lsim 10^{11-11.5} \lsun$. The model under-predicts the observed FIR LF at higher luminosities and at higher redshifts e.g. when comparing to ALMA REBELS results \citep{Barrufet23}. Even our maximal model, with $\feff=1.0$ and $f_{\rm{esc}}^{\rm{UV}}=0.0$ under-predicts the number density of the brightest objects, requiring further ALMA follow up and spectroscopic confirmations for these rare sources. 

\end{itemize}

Finally, we end with some caveats. Firstly, we find SNII to be the primary dust factories with ISM grain growth in a homogeneous medium playing a minor role. However, including a multi-phase ISM, with cold clumps where grain growth could be more efficient, could help increase the contribution of the latter process to the total dust mass. Secondly, we assume dust to be homogeneously distributed in the ISM. However, as has been shown by recent ALMA observations \citep[e.g.][]{Inami22}, dust and star forming regions can be spatially segregated, significantly affecting the dust optical depth experienced by UV photons. Thirdly, while we assume a Kroupa IMF throughout this work, the redshift evolution of the IMF remains an outstanding issue, for example becoming more top-heavy with decreasing metallicity \citep[see e.g][]{chon2021}. This could have a significant impact on the inferred UV luminosities  
\citep[e.g.][]{pacucci2022,yung2023}. Fourthly, we assume a constant star formation efficiency for massive systems, not accounting for observed galaxies lying significantly above the main sequence of star-formation \citep{harikane2022a, pacucci2022}. Finally, we have ignored the heating from the CMB which, in addition to setting the temperature floor for both the gas and dust temperatures, provides the background against which dust emission is observed. Forthcoming observations with JWST will be crucial in obtaining spectroscopic redshifts to validate the highest-redshift sources observed, with multi-band ALMA observations providing crucial constraints on the dust temperatures (and hence masses) of galaxies in the era of cosmic dawn.

\section*{Acknowledgments} 
VM and PD acknowledge support from the NWO grant 016.VIDI.189.162 (``ODIN"). PD warmly thanks the European Commission's and University of Groningen's CO-FUND Rosalind Franklin program. The authors thank L. Barrufet, J. Kerutt, L. Sommovigo and M. Trebitsch for their helpful comments and insightful discussions.

\section*{Data Availability}
Data generated in this research will be shared on reasonable request to the corresponding author.

\bibliographystyle{mnras}
\bibliography{Delphi_JWST}


\end{document}